\documentclass[aps,showpacs]{revtex4}
\usepackage{amssymb}
\usepackage[dvips]{graphicx}
\usepackage[english]{babel}
\usepackage{indentfirst}
\usepackage{amsxtra}
\usepackage{amsmath}
\usepackage{supertabular}
\usepackage{multirow}
\usepackage [mathcal]{eucal}
\newcommand{\beq}{\begin{equation}}
\newcommand{\eeq}{\end{equation}}
\newcommand{\beqn}{\begin{eqnarray}}
\newcommand{\eeqn}{\end{eqnarray}}
\newcommand{\sumint}{\sum \!\!\!\!\!\!\!\!\int }
\newcommand{\bsigma}{\mbox{\boldmath $\sigma$}}
\newcommand{\btau}{\mbox{\boldmath $\tau$}}
\newcommand{\half}{\frac{1}{2}}

\newcommand{\caf}{{\cal F}}
\newcommand{\cag}{{\cal G}}
\newcommand{\cah}{{\cal H}}
\newcommand{\cau}{{\cal U}}
\newcommand{\caw}{{\cal W}}

\newcommand{\ide}{{\mathbb I}}
\newcommand{\vint}{{\cal A}}
\newcommand{\br}{{\bf r}}

\newcommand{\oxy}{$^{16}$O~}
\newcommand{\caI}{$^{40}$Ca~}
\newcommand{\caII}{$^{48}$Ca~}

\newcommand{\pb}{$^{208}$Pb~}
\newcommand{\sixj}[6]{ \left\{ \begin{array}{ccc}
                               #1 & #2 & #3 \\
                                #4 & #5 & #6 
                               \end{array}
                        \right\} }

\begin{document}

\noindent
\title{Self-consistent 
Continuum Random Phase Approximation calculations
with finite-range interactions}

\author{V. De Donno, G. Co'}
\affiliation{Dipartimento di Fisica, Universit\`a del Salento and,
 INFN Sezione di Lecce, Via Arnesano, I-73100 Lecce, ITALY}
\author{M. Anguiano, A. M. Lallena}
\affiliation{Departamento de F\'\i sica At\'omica, Molecular y
  Nuclear, Universidad de Granada, E-18071 Granada, SPAIN}

\date{\today}

\bigskip

\begin{abstract}
We present a technique which treats, without approximations, the
continuum part of the excitation spectrum in Random Phase
Approximation calculations with finite-range interactions.  The
interaction used in the Hartree-Fock calculations to generate the
single particle basis is also used in the Continuum Random Phase
Approximation calculations. We show results for the electric dipole
and quadrupole excitations in the $^{16}$O, $^{22}$O, $^{24}$O,
$^{40}$Ca, $^{48}$Ca and $^{52}$Ca nuclei. We compare our results with
those of the traditional discrete Random Phase Approximation, with the
continuum independent particle model results and with those obtained
by a phenomenological Random Phase Approximation approach. We study
the relevance of the continuum, of the residual interaction and of the
self-consistency. We also compare our results with the available total
photoabsorption cross section data.  
\end{abstract}



\maketitle

\section{Introduction}
\label{sec:intro}
In the next few years, radioactive ion beams facilities will provide a
large amount of data on unstable nuclei. The description of the
structure of these nuclei is a challenge for the nuclear many-body
theories which have been mainly tested on stable nuclei.

The starting point of our description of nuclear systems is the
many-body Schr\"odinger equation with a two-body potential built to
describe elastic nucleon-nucleon (NN) cross sections and deuteron
properties \cite{wir95,mac01}.  To this two-body potential we add a
three-body force whose parameters are chosen to reproduce the $^3$H
binding energy \cite{pud97,pie02}.  Modern microscopic calculations,
which solve the many-body Schr\"odinger equation without
approximations, describe well the structure of nuclear few-body
systems as well as that of light nuclei \cite{pie01,kam01}.  These
results establish the validity of the non relativistic description of
atomic nuclei.  Unfortunately, the computational complexity of these
microscopic calculations hinders their application to medium and heavy
nuclei.

In recent years, there have been great advances in theories which solve
the many-body Schr\"odinger equation with microscopic interactions by
doing well controlled approximations
\cite{nav00,dea04,gan06,ari07,qua09,rot10}.  The results are very
promising but the calculations are still computationally rather
involved.  Probably, the development of these approaches will not be
rapid enough to cover the requirements for the description of the
data which will appear in the near future. We think that, for this
purpose, effective theories will play a fundamental role.

Effective theories search for solutions of the many-body Schr\"odinger
equation in a subspace of the full Hilbert space. Usually this
subspace is chosen to be formed by Slater determinants. This
restriction requires the modification of the NN
interaction in order to reproduce the energy eigenvalues of the
microscopic theory. By using Feshbach's projection techniques, it is
possible to obtain a formal expression which relates microscopic and
effective interactions \cite{rin80}. In the common practice, the
effective interaction is parametrized, and the values of the 
parameters are chosen to reproduce some experimental data. 

For example, in the J\"ulich approach \cite{spe77,spe80}, which we
shall call here phenomenological Random Phase Approximation (RPA)
approach, the single particle (s.p.) wave functions are produced by a
Woods-Saxon potential whose parameters are chosen to reproduce at best
charge radii and s.p. energies near the Fermi level of the nucleus
under investigation.  The values of the interaction parameters are
selected to reproduce the excitation energy of some particularly
collective state, for example, in the \pb nucleus, that of the
low-lying 3$^-$ state. The phenomenological approach has been applied
in the '80s of the past century to describe, and predict, with success
the excitation of the low-lying spectrum and of the giant resonances
mainly in doubly-magic nuclei. This phenomenological approach, based on
the Landau-Migdal theory of finite Fermi systems \cite{mig67},
requires the knowledge of a quite a number of observables in order to
select the RPA input parameters, i.e.  the s.p. basis and the
effective interaction. The philosophy of the approach requires to
change the input for every nucleus considered, therefore, despite of
its success, the phenomenological RPA approach is not suitable to
predict the structure of experimentally unknown nuclei. 

For this purpose, self-consistent RPA approaches are more useful. In
these approaches, s.p. wave functions and energies are obtained by
solving the Hartree-Fock (HF) equations. Since the effective
interaction used in HF is also used in the RPA calculation this type
of calculations is called self-consistent.  In the self-consistent
approaches the values of the parameters of the effective interaction
are chosen to reproduce binding energies, and charge radii, of a large
number of nuclei.  This fit produces an effective interaction to be
used in all the regions of the nuclear chart, including those so far
unexplored by the experiment.

Self-consistent RPA approaches have greater prediction power than
phenomenological approaches, but they require higher level of accuracy
and stability of the calculations. For example, the dimension of the
s.p. configuration space, beyond a certain size, is not a problem in
the phenomenological approach since the effects of the truncation of
the s.p. basis are taken into account by changing the values of the
interaction parameters. This procedure cannot be used any more in
self-consistent approaches, since the interaction parameters are
chosen once forever in HF calculations. This drawback of the
self-consistent RPA approach is avoided if the full s.p. configuration
space is used in the calculation. This implies a proper treatment of
the continuum part of the s.p. spectrum.

Self-consistent RPA approaches which consider the continuum have been
proposed already in the second half of the '70s
\cite{ber75,shl75,liu76,van81,war87,sar93,ham96}, but they are
applicable only if zero-range interactions are used.  In this case the
continuum RPA (CRPA) equations simplify, since direct and exchange
matrix elements assume the same expressions, i. e. they are
proportional.

Zero-range effective interactions have the great merit of simplifying
the calculations. There are however various drawbacks in their use,
many of them discussed already in Ref. \cite{dec80} where the D1
parameterization of the finite-range Gogny interaction was proposed.
Here we would like to mention some other aspects more directly related
to the present work. In RPA calculations zero-range interactions
produce more collectivity than finite-range interactions. The
difference becomes larger with the increasing value of the momentum
transfer \cite{mar07}. In addition, finite-range interactions provide
a better description of unnatural parity excitations
\cite{spe80,co90,don09}. Finally, finite-range interactions are more
directly comparable with microscopic NN force.

The reasons presented above motivated our work.  We developed a
formalism capable to solve the CRPA equations with finite-range
interactions. The main problem we had to face was the treatment of the
exchange terms of the interaction matrix elements.  In the literature
there are few examples of CRPA calculations done with finite-range
interactions \cite{uda89,bub91}, and, to the best of our knowledge,
only a single case of self-consistent CRPA calculation \cite{nak09}.
The relativistic self-consistent calculations of Ref. \cite{piek01}
use finite range interactions, but in the so-called Hartree
approximation where the exchange terms are neglected.  Our approach,
which will be described in Sect. \ref{sec:form}, follows the lines of
that proposed by the J\"ulich group at the beginning of the '90s
\cite{bub91}.  There are, however, important differences due to the
different manner of generating the s.p. bases. In the J\"ulich case
they are produced by using a Woods-Saxon potential while in our
approach with a HF calculation.

Details and basic ingredients of the calculations, such as
interactions, expansion basis, test of convergence etc., are presented
in Sect. \ref{sec:details}. A discussion of a selected set of results
is done in Sect. \ref{sec:results}.  We have calculated charge
conserving excitations in three oxygen isotopes and in three calcium
isotopes. We compare our CRPA results with those obtained from
discrete RPA calculations, with the results of the phenomenological
RPA approach and also with results obtained by switching off the
residual interaction in the RPA calculations. Following the notation
of Ref. \cite{boh81} we shall label independent particle model (IPM)
the results of this last type of calculations.  In
Sect. \ref{sec:conclusions} we summarize the main results of our work
and we draw our conclusions.

\section{Formalism}
\label{sec:form}
The RPA theory describes the excited state of a many-body system as
linear combination of particle-hole ($ph$), and hole-particle ($hp$)
excitations.  This implies the existence of a s.p. basis which, in our
calculations, is generated by solving the HF equations:
\beq 
\cah \, \left[ \phi_k(\br) \right] \, 
=\, -\frac{\hbar^2}{2m}\nabla^2\phi_k(\boldsymbol
r)\, + \, \cau(\boldsymbol r) \phi_k(\boldsymbol r) \, - \, \int {\rm
d}^3r' \, \caw(\boldsymbol r, \boldsymbol r')\, \phi_k(\boldsymbol r')
\, = \, \epsilon_k \, \phi_k(\br) \, ,
\label{eq:hf1}
\eeq 
where we have indicated with $\cah$ and $\phi_k$ the s.p. hamiltonian
and wave function respectively, with $\cau$ the Hartree potential
\beq 
\cau(\br)\, = \, \sum_{\alpha=1}^{8}\, \sum_{j \le F} \int {\rm d}^3 r'
\, \phi^*_j(\br') \, V_\alpha(\br,\br')\, \phi_j(\br') 
\, ,
\label{eq:hartree}
\eeq
with $\caw$ the Fock-Dirac potential  
\beq
\caw(\br, \br')\, = \, \sum_{\alpha=1}^{8} \, \sum_{j \le F} \,
   \phi^*_j(\br')\, V_\alpha(\br,\br')\, \phi_j(\br)  
\, ,
\label{eq:dirac}
\eeq
and with $\epsilon_k$ the energy eigenvalue of the $k$-th s.p. state.
In the above expressions, the sums on the s.p. states are restricted to
those with energy lower than the Fermi energy, i.e. to the hole
states. 

In our calculations we consider a two-body NN interaction of the form  
\beq
V_\alpha (\br_i,\br_j)= v_\alpha(|\br_i-\br_j|)\; O^\alpha_{i,j} 
\, , \,\,\, {\rm with} \,\,  
\alpha=1,2,\ldots,8 
\, ,
\label{eq:force1} 
\eeq
where $v_\alpha$ are scalar functions of the distance between the two
interacting nucleons, and $O^\alpha$ indicates the type of operator
dependence
\beqn
O^{\alpha}_{i,j}&:& 1\,,\,\,\btau(i)\cdot\btau(j)\,,\,\,
                 \bsigma(i)\cdot\bsigma(j)\,,\,\,
 \bsigma(i)\cdot\bsigma(j)\,\btau(i)\cdot\btau(j)\,, \nonumber
\\ &~&
S(i,j)\,,\,\, S(i,j) \btau(i)\cdot\btau(j)
\,,\,\, {\bf l}_{ij} \cdot {\bf s}_{ij}  
\,,\,\, {\bf l}_{ij} \cdot {\bf s}_{ij}\, \btau(i)\cdot\btau(j)  
\label{eq:fcahnnels}
\, .
\eeqn 
In the above expression we have indicated with 
$\bsigma$ the Pauli matrix operator acting on the spin
variable, with $\btau$ the analogous operator for the isospin, and
with 
\beq
S(i,j)\, = \,3 \, \frac {[\bsigma(i)\cdot (\br_i-\br_j)] \,
                    [\bsigma(j)\cdot (\br_i-\br_j)] } 
                 {(\br_i-\br_j)^2}\,
- \, \bsigma(i)\cdot\bsigma(j)
\eeq
the usual tensor operator. The terms $\alpha=7,8$ indicate the
spin-orbit contributions of the force. We include these last terms
only in the HF calculations and we consider them in a zero-range
approximation as it is done in Ref. \cite{dec80}.
 
We solve the HF equations in a spherical basis, and we express the
s.p. wave functions as
\beq
\phi^t_k(\br)\,=\, \phi^t_{nljm}(\br) \, = \, R^t_{nlj}(r)\,  
\sum_{s\mu} \, 
\langle l \mu \half s | j m \rangle \, Y_{l\mu}(\Omega) \, \chi_s  
\, .
\label{eq:spwf}
\eeq 
In the above equation, we have indicated with $n,l,j$ and $m$ the
principal quantum number, the orbital angular momentum, the total
angular momentum and its $z$-axis projection, respectively. We use $r$
to indicate the distance from the center of the nucleus, and $\Omega$
to indicate the usual angular coordinates of the polar spherical  
system.  The symbol $\langle \,\, |\,\rangle$ indicates the
Clebsch-Gordan coefficient, $Y_{l\mu}$ is the spherical harmonics and
$\chi_s$ the Pauli spinor for the spin. We call $t$ the third
component of the isospin and we use the convention $t=1/2$ for
protons and $t=-1/2$ for neutrons. The radial part of the s.p. wave
function satisfies the closure relation
\beq \sum_{\epsilon_i<\epsilon_{\rm F}} \, \delta_{ik} \, 
R_i(r)\, R^*_i(r')\, +\, \sumint_{\epsilon_k} \,
R_k(r,\epsilon_k) \, R^*_k(r',\epsilon_k) \, = \, \delta(r-r') 
\, ,
\label{eq:close} 
\eeq
where we have introduced the symbol \beq \sumint_{\epsilon_k} \,
\equiv \, \sum_{\epsilon_{\rm F} \le \epsilon_k \le 0} \, + \,
\int_0^\infty {\rm d} \epsilon_k \eeq
to indicate a sum and an integration on all the s.p. energies above
the Fermi surface. In the following, we shall indicate explicitly the
dependence on the s.p. energies $\epsilon_k$ which can assume both
discrete and continuum values. We shall use the index $h$ to indicate
all the quantum numbers identifying a hole s.p. state, energy
included, while the symbol $p$ will indicate all the quantum numbers
of the particle state, but the s.p. energy $\epsilon_p$.

A nuclear excited state $|\nu\rangle \equiv |J,\Pi,\omega \rangle$ is
characterized by its total angular momentum $J$, parity $\Pi$ and
excitation energy $\omega$. In the CRPA theory, the expression of the
operator that applied to the ground state generates the excited state
$|\nu\rangle$  can be expressed as
\beq
Q^\dag_\nu \, = \, \sum_{ph} \, \sumint_{\epsilon_p} \, 
\left[X^\nu_{ph}(\epsilon_p)\, a^\dag_p(\epsilon_p)\, a_h \,
     - \, Y^\nu_{ph}(\epsilon_p)\, a^\dag_h \, a_p(\epsilon_p) \right] \, ,
\label{eq:qnu}
\eeq
where we have indicated with $a^\dag$ and $a$ the usual
particle creation and annihilation operators and with $X$ and $Y$ the
RPA amplitudes.  The CRPA secular equations whose solution
provides the values of $X$ and $Y$ can be written as 
\beqn
(\epsilon_p-\epsilon_h-\omega)\, X_{ph}^\nu(\epsilon_p)\,+ \nonumber   
&& \\  & &
\hspace*{-3.5cm}
\sum_{p'h'} \, \sumint_{\epsilon_{p'}} \,
\left[v^J_{ph,p'h'}(\epsilon_p,\epsilon_{p'})\, X_{p'h'}^\nu(\epsilon_{p'})\,
   + \, u^J_{ph,p'h'}(\epsilon_p,\epsilon_{p'}) \, Y_{p'h'}^\nu(\epsilon_{p'}) 
     \right] \, = \, 0 \,,
\label{eq:rpa1}
\\
(\epsilon_p-\epsilon_h+\omega) \, Y_{ph}^{\nu}(\epsilon_{p})\,+ \nonumber 
&& \\ & &
\hspace*{-3.5cm}
\sum_{p'h'} \, \sumint_{\epsilon_{p'}}\, 
\left[
  v^{J*}_{ph,p'h'}(\epsilon_p,\epsilon_{p'})\, 
Y_{p'h'}^{\nu}(\epsilon_{p}) \,
 + \, u^{J*}_{ph,p'h'}(\epsilon_p,\epsilon_{p'}) \, 
X_{p'h'}^{\nu}(\epsilon_{p'})
   \right] \, = \, 0
\,.
\label{eq:rpa2}
\eeqn
In the above equations, the interaction terms have been defined as
\beqn
v^J_{ph,p'h'}(\epsilon_p,\epsilon_{p'}) &=& 
v^{J,{\rm dir}}_{ph,p'h'}(\epsilon_p,\epsilon_{p'}) \, - \,  
v^{J,{\rm exc}}_{ph,p'h'}(\epsilon_p,\epsilon_{p'}) 
\, , 
\label{eq:v}
\eeqn
and 
\beqn
u^J_{ph,p'h'}(\epsilon_p,\epsilon_{p'}) &=&
(-1)^{j_{p'} + j_{h'} - J} \, 
v^J_{ph,h'p'}(\epsilon_p,\epsilon_{p'}) 
\, ,
\label{eq:u}
\eeqn
with
\beqn 
&& \hspace*{-1cm}
v^{J,{\rm dir}}_{ph,p'h'}(\epsilon_p,\epsilon_{p'}) \, = \, \int
{\rm d}r_1\, r_1^2 \, \int {\rm d}r_2 \, r_2^2 \,
R^*_p(r_1,\epsilon_p) \, R^*_{h'}(r_2) \, V^{J,{\rm
dir}}_{ph,p'h'}(r_1,r_2) \, R_h(r_1) \, R_{p'}(r_2,\epsilon_{p'}) \, ,
\label{eq:vdir} \\ 
&& \hspace*{-1cm}
v^{J,{\rm exc}}_{ph,p'h'}(\epsilon_p,\epsilon_{p'})
\, =\,  \int {\rm d}r_1\, r_1^2 \, \int {\rm d}r_2 \, r_2^2 \,
R^*_p(r_1,\epsilon_p)\, R^*_{h'}(r_2) \, 
V^{J,{\rm exc}}_{ph,p'h'}(r_1,r_2) \, R_{p'}(r_1,\epsilon_{p'}) \, 
R_h(r_2) \, .
\label{eq:vexc}
\eeqn
In an analogous way, according to Eq. (\ref{eq:u}), we define the
corresponding $u$ functions. We have used the following definitions
for the quantities related to the interaction 
\beqn
V^{J,{\rm dir}}_{ph,p'h'}(r_1,r_2) &=& 
\sum_{\alpha=1}^{6} \, V_\alpha(r_1,r_2) \, 
\vint^{J,\alpha,{\rm dir}}_{ph,p'h'} 
\, , \label{eq:vbbdir} 
\\
V^{J,{\rm exc}}_{ph,p'h'}(r_1,r_2) &=& 
\sum_{\alpha=1}^{6} \, V_\alpha(r_1,r_2)  \,
\vint^{J,\alpha,{\rm exc}}_{ph,p'h'} \, , \label{eq:vbbexc} 
\\
\vint^{J,\alpha,{\rm dir}}_{ph,p'h'} &=&
\sum_K (-1)^{j_h + j_{p'} + K} \, \hat{K} \, 
\sixj {j_p} {j_h} {J}{j_{p'}}{j_{h'}}{K} \,
\langle j_p j_{h'} K \| V_\alpha(\Omega) \| j_h j_{p'} K\rangle \,, 
\\
\vint^{J,\alpha,{\rm exc}}_{ph,p'h'} &=&
\sum_K  \, \hat{K} \, 
\sixj {j_p} {j_h} {J}{j_{p'}}{j_{h'}}{K} \,
\langle j_p j_{h'} K \| V_\alpha(\Omega) \|  j_{p'} j_h K\rangle \, .
\eeqn
Here for the angular momentum quantum numbers we used the notation
$\hat{a}~=~\sqrt{2 a + 1}$.  The terms in curly brackets are the Racah
$6j$ symbol and the double bars indicate the reduced matrix element as
defined by the Wigner-Eckart theorem. In our work we adopt the phase
conventions of Ref. \cite{edm57}.

In the above equations we have factorized the two-body NN interaction
(\ref{eq:force1}) in a radial part $V_\alpha(r_1,r_2)$, depending only on
the moduli of the positions of the two-interacting nucleons, and in an
angular and operator dependent part $V_\alpha(\Omega)$. We have done this
factorization by using the Fourier transformed expression of the NN
interaction.

Our method of solving the CRPA equations consists in reformulating the
secular equations (\ref{eq:rpa1}) and (\ref{eq:rpa2}) with new unknown
functions which do not have an explicit dependence on the continuous
particle energy $\epsilon_p$. This is the same approach adopted in
Refs. \cite{uda89,bub91}, but in our case the s.p. wave functions are
generated by a HF calculation. For this reason, we present here, with
some detail, the various steps bringing to the new CRPA secular
equations. The new unknowns are the channel functions $f$ and $g$
defined as:
\beq 
f^\nu_{ph}(r) \, = \, \sumint_{\epsilon_p} \, 
X^\nu_{ph}(\epsilon_p) \, R_p(r,\epsilon_p) \, ,
\label{eq:f}
\eeq 
and
\beq 
g^\nu_{ph}(r) \, = \, \sumint_{\epsilon_p} \, 
Y^{\nu*}_{ph}(\epsilon_p) \, R_p(r,\epsilon_p)  \, .
\label{eq:g}
\eeq 

The first step of this procedure consists in multiplying Eqs.
(\ref{eq:rpa1}) and (\ref{eq:rpa2}) by $ R_p(r,\epsilon_p)$, the
radial part of the particle wave function. Considering
Eq. (\ref{eq:hf1}), we obtain for the left hand side of Eq. (\ref{eq:rpa1})
\beq (\epsilon_p-\epsilon_h-\omega)\, 
R_p(r,\epsilon_p)\,X_{ph}^\nu(\epsilon_p) \, = \,
\cah \,\left[R_p(r,\epsilon_p) \, X_{ph}^\nu(\epsilon_p) \right]   \,  - \,
(\epsilon_h+\omega) \, R_p(r,\epsilon_p) \, X_{ph}^\nu(\epsilon_p) \, .  
\eeq
The second step of the procedure is to integrate on $\epsilon_p$, the
particle energy. For the first term in the right hand side of the
above equation we obtain, using again Eq. (\ref{eq:hf1}),
\beqn 
\nonumber 
\sumint_{\epsilon_p}\, 
\cah \,\left[ R_p(r,\epsilon_p) \, X_{ph}^\nu(\epsilon_p)\right] 
& =& \nonumber
-\, \frac{\hbar^2}{2m}\, \nabla^2 \, \sumint_{\epsilon_p} \,
R_p(r,\epsilon_p)\, X_{ph}^\nu(\epsilon_p) \, 
+ \, \cau(\br) \, \sumint_{\epsilon_p} \, R_p(r,\epsilon_p)
\, X_{ph}^\nu(\epsilon_p) \\ \nonumber 
&&\hspace*{0.5cm}- \, \
\int {\rm d}^3r' \, \caw(\boldsymbol r, \boldsymbol r')\, 
\sumint_{\epsilon_p}\, R_p(r,\epsilon_p)\, X_{ph}^\nu(\epsilon_p) 
\\ \nonumber
&=& -\, \frac{\hbar^2}{2m}\, \nabla^2 \, f^\nu_{ph}(r)\,  +\, 
\cau(\br)\, f^\nu_{ph}(r)\, -\, \int
{\rm d}^3r' \, \caw(\br,\br') \, f^\nu_{ph}(r') 
\\ &=& \cah \, \left[ f^\nu_{ph}(r) \right] 
\, .
\eeqn 
In the above equation we have indicated with $\nabla^2$ the usual
Laplace operator where the differential terms related to $\theta$ and
$\phi$ have been already applied to the spherical harmonics providing
the correct eigenvalue.  Therefore, in our writing, we imply that only
the derivations on $r$ should be done.

The operations described above are applied to all the terms of
Eqs. (\ref{eq:rpa1}) and (\ref{eq:rpa2}).  As example of our
calculations, we write the contribution of the first interaction term
of Eq.~(\ref{eq:rpa1}): 
\beqn 
\sumint_{\epsilon_p} \, R_p(r,\epsilon_p) \, \sum_{p'h'}\,
\sumint_{\epsilon_{p'}} \, v^J_{ph,p'h'}(\epsilon_p,\epsilon_{p'}) \,
X^\nu_{p'h'}(\epsilon_p') &=& \nonumber \\ 
\nonumber && \hspace*{-7.5cm} = \,
\sumint_{\epsilon_p} \, R_p(r,\epsilon_p) \, \sum_{p'h'} \,
\sumint_{\epsilon_{p'}} \, 
\int {\rm d}r_1\, r_1^2 \, \int {\rm d}r_2 \, r_2^2 \, 
R^*_p(r_1,\epsilon_p) \, R^*_{h'}(r_2)
\\ 
\nonumber &~& 
\hspace*{-7cm}\Bigg[ V^{J,{\rm dir}}_{ph,p'h'}(r_1,r_2) \, 
R_h(r_1) \, R_{p'}(r_2,\epsilon_{p'}) \, - \, 
V^{J,{\rm exc}}_{ph,p'h'}(r_1,r_2)
\, R_{p'}(r_1,\epsilon_{p'}) \, R_h (r_2) \Bigg] \, 
X^\nu_{p'h'}(\epsilon_{p'}) \nonumber
\\ 
\nonumber && \hspace*{-7.5cm} = \,
\sumint_{\epsilon_p} \, R_p(r,\epsilon_p) \, 
\sum_{p'h'} \, \int {\rm d}r_1\, r_1^2 \, 
\int {\rm d}r_2 \, r_2^2 \, R^*_p(r_1,\epsilon_p) \, R^*_{h'}(r_2) \, 
\\ 
\nonumber &~& 
\hspace*{-7cm}\Bigg[ V^{J,{\rm dir}}_{ph,p'h'}(r_1,r_2) \, 
R_h(r_1) \,  f^\nu_{p'h'}(r_2) \, - \, V^{J,{\rm exc}}_{ph,p'h'}(r,r_2)
\, f^\nu_{p'h'}(r_1) \, R_h (r_2) \Bigg] \nonumber
\\ 
\nonumber && \hspace*{-7.5cm} = \,
\sum_{p'h'} \, \int {\rm d}r_2 \, r_2^2 \, R^*_{h'}(r_2) 
\\
&~& 
\hspace*{-7cm}\Bigg[
  V^{J,{\rm dir}}_{ph,p'h'}(r,r_2)\, R_h(r) \, f^\nu_{p'h'}(r_2)
- V^{J,{\rm exc}}_{ph,p'h'}(r,r_2)\, f^\nu_{p'h'}(r) \, R_h (r_2) \Bigg]\,
+ \, {\cal T}(r)
\,,
\eeqn
In the above equation we have used the definition (\ref{eq:f}) and the
closure relation (\ref{eq:close}) and we have defined the term
\beqn
{\cal T}(r)& =&  - \, \sum_{\epsilon_i<\epsilon_{\rm F}} \, 
\delta_{ip} \, R_i(r) \, \int {\rm d}r_1 \, r_1^2 \, R^*_i(r_1) 
\, \int {\rm d}r_2 \, r_2^2 \, \sum_{p'h'}  \, R^*_{h'}(r_2)
\nonumber \\ &~& 
\Bigg[ V^{J,{\rm dir}}_{ph,p'h'}(r_1,r_2)\, R_h(r_1) \, f_{p'h'}(r_2)
- V^{J,{\rm exc}}_{ph,p'h'}(r_1,r_2)\, f_{p'h'}(r_1) \, R_h (r_2)
\Bigg] 
\, ,
\eeqn
where, to simplify the writing, we have dropped the dependence of $f$
on $\nu \equiv (J,\Pi,\omega)$.

We write a new set of CRPA secular equations where the unknowns  
are the channel functions $f$ and $g$,  
\beqn 
&~&
\hspace*{-1.5cm}
\cah \, \left[ f_{ph}(r) \right]
 -\, (\epsilon_h\, +\, \omega)\, f_{ph}(r) \,=\, 
-\, \caf^J_{ph}(r) \, +
\, \sum_{\epsilon_i<\epsilon_{\rm F}} \, \delta_{ip} \, R_i(r) \, 
\int {\rm d}r_1 \, r_1^2 \, R^*_i(r_1) \, \caf^J_{ph}(r_1) \, ,
\label{eq:feq}\\
&~&
\hspace*{-1.5cm}\cah \, \left[ f_{ph}(r) \right] -
(\epsilon_h\, -\, \omega)\, g_{ph}(r) \, =\,
-\, \cag^J_{ph}(r) \, +
\, \sum_{\epsilon_i<\epsilon_{\rm F}} \, \delta_{ip} \, R_i(r) \, 
\int {\rm d}r_1 \, r_1^2 \, R^*_i(r_1) \, \cag^J_{ph}(r_1) \, ,
\label{eq:geq}
\eeqn
where we have defined 
\beqn
\hspace*{-1.cm}
\caf^J_{ph}(r) \, = \, \sum_{p'h'} \, \int {\rm d}r_2 \, r_2^2 
&& \nonumber \\  
&& \nonumber \hspace*{-3.cm}
\Bigg\{ R^*_{h'}(r_2) \, \Bigg[
V^{J,{\rm dir}}_{ph,p'h'}(r,r_2)\, R_h(r) \, f_{p'h'}(r_2)
- V^{J,{\rm exc}}_{ph,p'h'}(r,r_2)\, f_{p'h'}(r) \, R_h (r_2) \Bigg]
\\
&&  \hspace*{-3.cm} + \, g^*_{p'h'}(r_2) \, 
\Bigg[ U^{J,{\rm dir}}_{ph,p'h'}(r,r_2) \, R_h(r)\, R_{h'}(r_2)\, -  
 \, U^{J,{\rm exc}}_{ph,p'h'}(r,r_2 ) \, R_{h'}(r) \, R_h(r_2) \Bigg]  
\Bigg\} 
\, ,
\label{eq:Fcal}
\eeqn
and $\cag^J_{ph}$ is obtained from the above equation by interchanging
the $f$ and $g$ channel functions.  The relation between the $U$ and
$V$ symbols is analogous to that of the $u$ and $v$ symbols of
Eq. (\ref{eq:u}).  The last terms in the right-hand side of
Eqs. (\ref{eq:feq}) and (\ref{eq:geq}) are zero if there are no hole
states having the same angular momenta $l$ and $j$ of the particle
state considered. 

We have changed a set of algebraic equations with unknowns depending
on the continuous variable $\epsilon_p$ into a set of
integro-differential equations with unknowns depending on the 
distance from the center of coordinates.  The solution of this problem
requires to impose the proper boundary conditions. If the excitation
energy $\omega$ is above the nucleon emission threshold, in some of
the $ph$ excitation pairs compatible with the angular momentum and the
parity of the final state, the particle has positive energy. We call
open channels these $ph$ pairs, and closed channels those pairs where
the particle is in a discrete state.

After fixing the angular momentum $J$ and the parity $\Pi$ of the
excited state, for each value of the excitation energy $\omega$, we
solve Eqs. (\ref{eq:feq}) and (\ref{eq:geq}) a number of times equal
to the number of the open channels. Every time we impose a different
boundary condition, i.e. that the particle is emitted only in a
specific channel, which we call elastic channel and we label it as
$p_0h_0$. For an open $ph$ channel, we impose that the outgoing
asymptotic behaviour of the channel function $f_{ph}^{p_0 h_0}$ is 
\beq
f_{ph}^{p_0 h_0}(r\to\infty)\, \to \,
R_{p_0}(r,\epsilon_p)\, \delta_{p,p_0}\, \delta_{h,h_0}\, +\, \lambda
\, H^-_p(\epsilon_h+\omega,r)
\,,
\label{eq:asintf} 
\eeq 
where $\lambda$ is a complex normalization constant and
$H^-_p(\epsilon_h+\omega,r)$ is an ingoing Coulomb function or a
Hankel function in case of a proton or neutron channel,
respectively. The s.p. wave function $R_{p}$ is eigenfunction of the
HF hamiltonian (\ref{eq:hf1}) for positive energy, and is
calculated as described in Appendix \ref{app:cwf}.

In the case of a closed channel, the asymptotic behaviour is given by 
a decreasing exponential function 
\beq
f_{ph}^{p_0 h_0}(r\to\infty) \, 
\rightarrow \, 
\frac{1}{r}\,
\exp\left[-r\left(\frac{2m|\epsilon_h+\omega|}{\hbar^2}\right)^{\frac{1}{2}}
\right]
\,,
\label{eq:asintf1}
\eeq
as in the case of the channel functions
$g_{ph}^{p_0 h_0}$, 
\beq
g_{ph}^{p_0 h_0}(r\to\infty)
\, \rightarrow \,
\frac{1}{r} \,
\exp\left[-r\left(\frac{2m|\epsilon_h-\omega|}{\hbar^2}\right)^{\frac{1}{2}}
\right]
\,.
\label{eq:asintg}
\eeq

We solve the CRPA secular equations (\ref{eq:feq}) and (\ref{eq:geq})
by using a procedure similar to that presented in Ref. \cite{bub91}.
The channel functions $f$ and $g$ are expanded on a basis of sturmian
functions $\Phi^{\mu}_p$ which obey the required boundary conditions
(\ref{eq:asintf})-(\ref{eq:asintg}).

The sturmian functions $\Phi^{\mu}_p$ are defined as eigenstates of the
differential equation \cite{rot62, wei63, wei64, new66, raw82}
\beq
\left[-\,\frac{\hbar^2}{2m}\,\frac{{\rm d}^2}{{\rm d}r^2} \,
- \, \frac{\hbar^2}{m}\, \frac{1}{r}\, \frac{\rm d}{{\rm d}r}\,
+ \, \frac{\hbar^2}{2m}\, \frac{l_p(l_p+1)}{r^2}\ -\, \epsilon_p\right]
\, \Phi^\mu_p(r) \ =\,  -\,\overline{U}_p^\mu(r) \, \Phi^\mu_p(r)
\, ,
\label{eq:sturmb}
\eeq
where $m$ is the particle mass, $l_p$ is the orbital quantum number and
$\overline U_p^\mu(r)$ is a complex square well potential of the form
\beq
\overline U_{p}^{\mu}(r) \
=\, \left\{
 \begin{array}{ll} 
  \beta^{\mu}_p+i \gamma_{p}^{\mu} \, ,
 &\mbox{if $r\le a$} \, ,
\\
0 \, , &\mbox{if $r>a$} \, ,
\end{array}
\right.
\label{eq:potU}
\eeq 
with $\beta^{\mu}_p$ and $\gamma^{\mu}_p$ real constants.  The
requirement of continuity of $\Phi^{\mu}_p$ at $r=a$ implies that
only a discrete set of values of $\beta^{\mu}_p$ and
$\gamma^{\mu}_p$ should be considered. In this set of solutions, the
index $\mu$ is related to the number of nodes of the Sturm-Bessel
function $\Phi^{\mu}_p$ in the region $0 \le r \le a$. When the
value of the index $\mu$ increases by one unity, an additional node
appears in the wave function at $r \le a$.  The definition of the
Sturm-Bessel functions given above implies the orthogonality relation 
\beq
(\beta^\mu_p+i\gamma^\mu_p)\, 
\int_0^{a} {\rm d}r \, r^2 \, \Phi^\mu_p(r) \, \Phi^\nu_p(r) \,
= \, \delta_{\mu\nu}
\label{eq:ort1}
\, .  
\eeq

Since, in general, the Sturm-Bessel functions are not
orthogonal to the wave functions of the s.p. hole states, we find
more useful to consider a set of orthogonalized functions which we
construct as
\beq
\widetilde{\Phi}^\mu_{p}(r) \,= \, 
\Phi^\mu_{p}(r)\, - \, \sum_{\epsilon_i<\epsilon_{\rm F}} \, \delta_{ip} \,
R^*_i(r) \, \int {\rm d}r' \, r'^2 \, R_i(r') \, \Phi^\mu_p(r')
\, ,
\label{eq:orthost}
\eeq
where with 
$\delta_{ip}$ we 
indicate that in the sum $l_i=l_p$ and $j_i=j_p$.
By using this set of
orthogonalized sturmian functions we express the channel functions
$f_{ph}^{p_0h_0}$ and $g_{ph}^{p_0h_0}$ as
\beqn
f_{ph}^{p_0h_0}(r)&=&R_{p_0}(r,\epsilon_{p_0})\, \delta_{p p_0}\, \delta_{hh_0} \, 
+ \, \sum_\mu \, c^{\mu+}_{ph} \, \widetilde{\Phi}^{\mu+}_{p}(r) 
\, ,
\label{eq:expf}
\\
g_{ph}^{p_0h_0}(r)&=&
\sum_\mu \, c^{\mu-}_{ph} \, \widetilde{\Phi}^{\mu-}_{p}(r)
\, ,
\label{eq:expg}
\eeqn
where the superscripts $+$ and $-$ indicate that the sturmian
functions are calculated for $\epsilon_p=\epsilon_h+\omega$ or
$\epsilon_p=\epsilon_h-\omega$ respectively. To simplify the writing
we drop the explicit dependence on the open channel label $p_0h_0$ of 
all the $c^\mu_{ph}$ expansion coefficients.

We insert the expressions (\ref{eq:expf}) and (\ref{eq:expg}) in the
secular equations (\ref{eq:feq}) and (\ref{eq:geq}), and following the
steps presented in Appendix \ref{app:sexp}, we obtain a system of
linear equations whose unknowns are the expansion coefficients
$c^{\mu\pm}_{ph}$. The new CRPA secular equations are
\beqn
\hspace*{-0.5cm} 
\sum_\mu \, \sum_{p'h'} \, \Bigg\{ \Bigg[ \delta_{pp'} \, \delta_{hh'}
\Big( \delta_{\mu \nu} \, -
\, \langle (\Phi_p^{\nu +})^*|\cau|\Phi_p^{\mu+} \rangle \,+ \, 
\langle (\Phi_p^{\nu +})^*\, \ide |\caw| \ide \,\Phi^{\mu+}_p \rangle \nonumber \\
&& \hspace*{-7.6cm} + \, \sum_{\epsilon_i<\epsilon_{\rm F}}\, \delta_{ip}\, (\epsilon_i-\epsilon_h-\omega)\,
\langle (\Phi_p^{\nu +})^* |R_i\rangle \langle
(R_i)^*|\Phi_p^{\mu+}\rangle \Big)\nonumber \\
&& \hspace*{-10.cm}  - \, \Big(
\langle (\widetilde{\Phi}_p^{\nu +})^* R_{h'} |  
V^{J,{\rm dir}}_{ph,p'h'} |R_h \widetilde{\Phi}^{\mu+}_{p'} \rangle \,
- \, \langle (\widetilde{\Phi}_p^{\nu +})^* R_{h'} |
   V^{J,{\rm exc}}_{ph,p'h'} | \widetilde{\Phi}^{\mu+}_{p'} R_h  \rangle \Big)
\Bigg] \, c^{\mu+}_{p'h'} \, \nonumber\\
&& \hspace*{-10.cm}  - \, \Big(
\langle (\widetilde{\Phi}_p^{\nu +})^*  \widetilde{\Phi}^{\mu-}_{p'}|  
U^{J,{\rm dir}}_{ph,p'h'} | R_h R_{h'} \rangle \, 
- \, \langle (\widetilde{\Phi}_p^{\nu +})^*  \widetilde{\Phi}^{\mu-}_{p'}|  
U^{J,{\rm exc}}_{ph,p'h'} | R_{h'} R_h \rangle \Big) \, (c^{\mu-}_{p'h'})^*
\Bigg\} \, = \, \nonumber \\
&& \hspace*{-10.5cm}
= \, \langle (\widetilde{\Phi}_p^{\nu +})^* R_{h_0} |
   V^{J,{\rm dir}}_{ph,p_0h_0} |R_h R_{p_0}(\epsilon_{p_0}) \rangle \, 
- \, \langle (\widetilde{\Phi}_p^{\nu +})^* R_{h_0} |
   V^{J,{\rm exc}}_{ph,p_0h_0} |R_{p_0}(\epsilon_{p_0}) R_h  \rangle 
\, ,
\label{eq:st1} 
\eeqn
\beqn
\hspace*{-0.5cm} 
\sum_\mu \, \sum_{p'h'} \, \Bigg\{ \Bigg[ \delta_{pp'} \, \delta_{hh'}
\Big( \delta_{\mu \nu} \, -
\, \langle (\Phi_p^{\nu -})^*|\cau|\Phi_p^{\mu-} \rangle \,+ \, 
\langle (\Phi_p^{\nu -})^*\, \ide |\caw| \ide \,\Phi^{\mu-}_p \rangle \nonumber \\
&& \hspace*{-7.6cm} + \, \sum_{\epsilon_i<\epsilon_{\rm F}}\, \delta_{ip}\, (\epsilon_i-\epsilon_h+\omega)\,
\langle (\Phi_p^{\nu -})^* |R_i\rangle \langle
(R_i)^*|\Phi_p^{\mu-}\rangle \Big)\nonumber \\
&& \hspace*{-10.cm}  - \, \Big(
\langle (\widetilde{\Phi}_p^{\nu -})^* R_{h'} |  
V^{J,{\rm dir}}_{ph,p'h'} |R_h \widetilde{\Phi}^{\mu-}_{p'} \rangle \,
- \, \langle (\widetilde{\Phi}_p^{\nu -})^* R_{h'} |
   V^{J,{\rm exc}}_{ph,p'h'} | \widetilde{\Phi}^{\mu-}_{p'} R_h  \rangle \Big)
\Bigg] \, c^{\mu-}_{p'h'} \, \nonumber\\
&& \hspace*{-10.cm}  - \, \Big(
\langle (\widetilde{\Phi}_p^{\nu -})^*  \widetilde{\Phi}^{\mu+}_{p'}|  
U^{J,{\rm dir}}_{ph,p'h'} | R_h R_{h'} \rangle \, 
- \, \langle (\widetilde{\Phi}_p^{\nu -})^*  \widetilde{\Phi}^{\mu+}_{p'}|  
U^{J,{\rm exc}}_{ph,p'h'} | R_{h'} R_h \rangle \Big) \, (c^{\mu+}_{p'h'})^*
\Bigg\} \, = \, \nonumber \\
&& \hspace*{-10.5cm}
= \, \langle (\widetilde{\Phi}_p^{\nu -})^* R_{p_0}(\epsilon_{p_0}) |
   U^{J,{\rm dir}}_{ph,p_0h_0} |R_h R_{h_0} \rangle \, 
- \, \langle (\widetilde{\Phi}_p^{\nu -})^* R_{p_0}(\epsilon_{p_0}) |
   U^{J,{\rm exc}}_{ph,p_0h_0} |R_{h_0} R_h  \rangle \, .
\label{eq:st2} 
\eeqn

In the above expressions, with the bra and ket integration convention
we indicate integrations on radial variables only.  The number of
these integrations is given by the number of the functions inserted
between the bra and ket symbols. For this reason we have inserted the
symbol $\ide$ indicating the identity function.

Summarizing, we have converted the CRPA secular equations
(\ref{eq:rpa1}) and (\ref{eq:rpa2}) into a set of algebraical
equations whose unknowns are the expansion coefficients $c^\mu_{ph}$.
These equations have a solution for each value of the excitation
energy $\omega$ above the nucleon emission threshold.

The solution of the secular CRPA equations provides the channel
functions $f$ and $g$ and this allows us to calculate the transition
matrix elements induced by an operator $T_J$. If the operator $T_J$
inducing the transition is of one-body type of the form 
\beqn
T_{JM}(\br)\, = \, \sum_{i=1}^A \, F_J(r_i) \, \theta_{JM}(\Omega_i)
\, \delta(\br_i-\br)
\, ,
\eeqn
where we have separated the dependence on the radial and angular parts
of the operator, we obtain for the transition matrix element the
expression 
\beqn
\langle J \| T_J \| 0 \rangle_{p_0h_0} &=&  \, 
\sum_{p h}\,  
\left[ \langle j_p \| \theta_J \| j_h \rangle \,  
\int {\rm d}r \, r^2 \, (f^{p_0h_0}_{ph}(r))^* \, F_J(r) \, R_h(r)  
\right.\nonumber
\\
&& \hspace*{1.2cm} \left. + \, (-1)^{J+j_p -j_h} \, 
\langle j_h \| \theta_J \| j_p \rangle \,  
\int {\rm d}r \, r^2 \,  R^*_h(r) \, F_J(r) \, g^{p_0h_0}_{ph}(r) \right] 
\, ,
\label{eq:transs1}
\eeqn 
where with the double bar we indicate the reduced matrix
elements of the angular coordinates, as defined in \cite{edm57}. 

In this paper we present results regarding nuclear excitations induced
by photons. We consider here only natural parity, electric,
excitations, and we use the following expression for the operator
$T_J$ 
\beq
T_{JM} \, = \,
\sum_{i=1}^A \, Z_{i}^{\rm eff} \, r_i^J \, Y_{JM}(\Omega_i) \, 
\delta(\br_i-\br) 
\, ,
\eeq
where $Z_{i}^{\rm eff}$ is the effective charge 
\beq
 Z_{i}^{\rm eff} \, = \, 
\left\{ 
\begin{array}{ll} 
\displaystyle 
\left( \frac{N}{A} \right)\, \half \, [1+\tau_3(i)] \,
- \, \left( \frac{Z}{A} \right) \, \half \, [1-\tau_3(i)] \, ,
          & \:\:\:  \mbox{if $J^\Pi=1^-$} \\
 & \\
\displaystyle  \half \, [1+\tau_3(i)] \, ,   & \:\:\:  \mbox{otherwise} \,\,.\\
\end{array}
\right. 
\eeq
The second of these expressions is obtained by using an approximation 
valid for the medium-heavy nuclei we are studying \cite{eisII}. 
In the above equation $A$, $Z$ and $N$ are the mass, proton and neutron
numbers respectively, and $\tau_3(i)=1$ for protons and
$-1$ for neutrons. 
For a given excitation energy $\omega$, and electric transition $EJ$,
we calculate the $B$-value as the incoherent sum on every open channel
$p_0h_0$,
\beq 
B(\omega,EJ:0 \to J)\, =  
\sum_{p_0h_0} |\langle \omega,J \| T_J \| 0\rangle_{p_0h_0}|^2 
\, .
\label{eq:bv} 
\eeq
We obtain the total photoabsorption cross section 
from the $B$-value by using the expression \cite{bla52}
\beqn 
\sigma(\omega,0 \to J)=
8\, \pi^3 \, \frac{J+1}{J} \, \frac{e^2}{[(2J+1)!!]^2} \left( 
\frac{\omega} {\hbar c}\right)^{2J-1} \, B(\omega,EJ:0\to J) 
\, ,
\label{eq:crossph}
\eeqn
where we have indicated with $e$ the elementary charge. 


\section{Details of the calculations}
\label{sec:details}
The formalism developed in the previous section leads to a set of
algebraic equations whose unknowns are the expansion coefficients
$c^{\mu \pm}_{ph}$. The number of coefficients, and therefore the
dimensions of the complex matrix to diagonalize, is an input of our
approach.

Since the expansion on a basis of sturmian functions is a technical
artifact, the solution of the CRPA secular equations must be
independent of the number of expansion coefficients. We tested the
convergence of our results by controlling the values of the total
photoabsorption cross section in \oxy and \caI nuclei. We reached the
stability up to the fifth significant figure with 10 expansion
coefficients, independently of the multipolarity and of the energy of the
excitation.

In our calculations we have used two different parameterizations of
the Gogny interaction, the more traditional D1S force \cite{ber91} and
the new D1M force \cite{gor09} obtained from a fit to about 2000
nuclear binding energies and 700 charge radii. The D1S and D1M forces
describe the empirical saturation point of symmetric nuclear matter
and reproduce rather well the behaviour of the equations of state
calculated with microscopic approaches \cite{akm98,gan10}. The
situation for pure neutron matter is different, because the behaviour
of the D1S equation of state at high densities is unphysical. The D1M
force produces an equation of state which has a plausible behaviour at
higher densities, even though it does not reproduce the results of
modern microscopic calculations.

\begin{figure}[ht]
\begin{center}
\includegraphics [scale=0.4,angle=90]{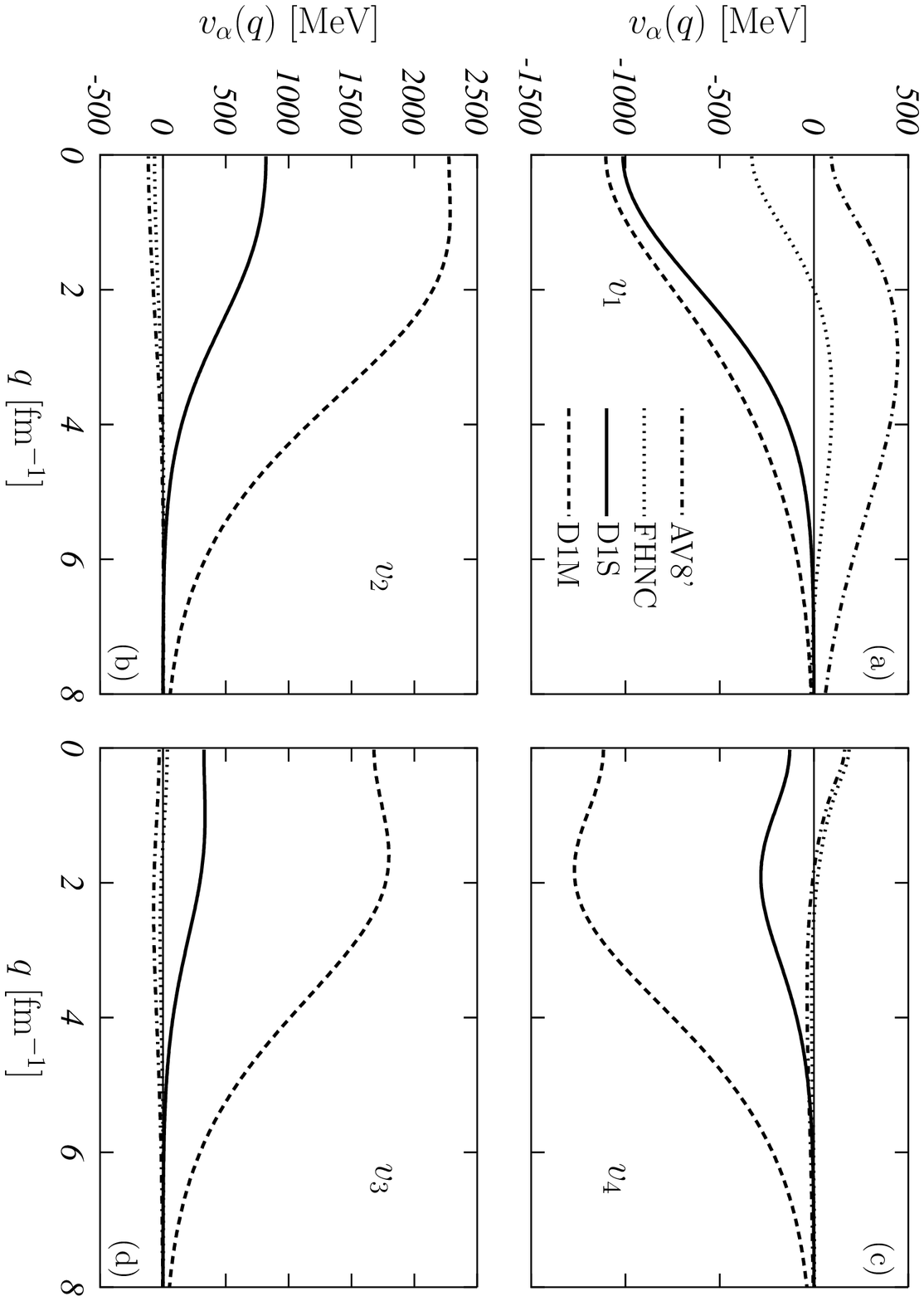}
\caption{\small Comparison between the interactions used in our
  calculations and the microscopic Argonne V8' interaction in the
  scalar channel, $v_1$, spin channel, $v_2$, isospin channel, $v_3$,
  and in the spin-isospin channel, $v_4$. 
  The dashed-dotted lines, labelled FHNC, have been obtained
  by multiplying the Argonne V8' terms with the scalar part of the 
  correlation function obtained in Correlated Basis Function
  calculations \cite{ari07}. 
}
\label{fig:force}
\end{center}
\end{figure}

In Fig. \ref{fig:force} we show the D1S and D1M forces in momentum
space, and we compare them with the bare Argonne V8' interaction
\cite{pud97}.  The scalar, $v_1$ (panel (a)), isospin, $v_2$ (panel
(b)), spin, $v_3$ (panel (c)) and spin-isospin, $v_4$ (panel (d)),
channels are shown. The dashed-dotted lines have been obtained by
multiplying the Argonne V8' terms with the scalar part of the
correlation function obtained in the finite nuclei Correlated Basis
Function calculations of Ref. \cite{ari07}. To be precise we have used
the correlation function obtained for the \caII nucleus. In any case,
these correlation functions are rather similar for all the nuclei
considered (see Fig. 21 of Ref. \cite{ari07}).  The large differences
between microscopic and effective interactions shown in
Fig. \ref{fig:force} indicate that the effective interactions take
into account a large number of effects explicitly treated in
microscopic calculations. The effects originated by the short-range
correlations are only a limited part of them.  It is interesting to
notice, in the spin, isospin and spin-isospin channels, the extremely
large values of the D1M interaction, with respect to those of the
other forces.  

In the next section we shall compare self-consistent CRPA results with
those of phenomenological calculations. The s.p. wave functions of the
phenomenological calculations have been obtained by using Woods-Saxon
wells, whose parameters are given in Ref. \cite{ari07}. The residual
interaction is a zero-range, density dependent, Landau-Migdal force
whose parameters are those chosen in Ref. \cite{co09b}.  

We have investigated nuclei where the hole s.p. levels are fully
occupied. This eliminates deformations and minimizes pairing effects.

\section{Results}
\label{sec:results}

\subsection{Oxygen}
\label{sec:oxygen} 
\begin{figure}[ht]
\begin{center}
\includegraphics[scale=0.4, angle=0] {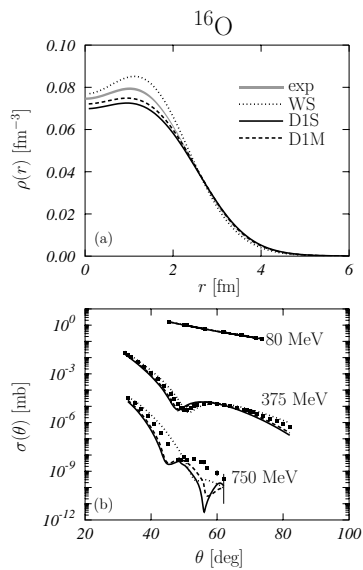} 
\caption{\small Panel (a): charge density distributions of the \oxy
  nucleus. The shaded area represents the empirical density
  distribution \cite{dej87}.  The other lines show the distributions
  obtained in HF calculations with the D1S and D1M interactions (full
  and dashed lines, respectively) and that obtained with the
  Woods-Saxon potential of Ref. \cite{ari07} (dotted line). Panel
  (b): elastic electron scattering cross sections calculated in
  Distorted Wave Born Approximation by using the charge distributions
  shown in the upper panel, as a function of the scattering angle
  $\theta$.  
  The empirical charge distribution plotted
  in the upper panel (dashed-dotted line) has been obtained from a fit
  to the data taken from Ref. \cite{sic70,sic}. 
  The numbers in the panel indicate the values of the electron energy.
  }
\label{fig:eeox}
\end{center}
\end{figure}

\begin{figure}[ht]
\begin{center}
\includegraphics[scale=0.4] {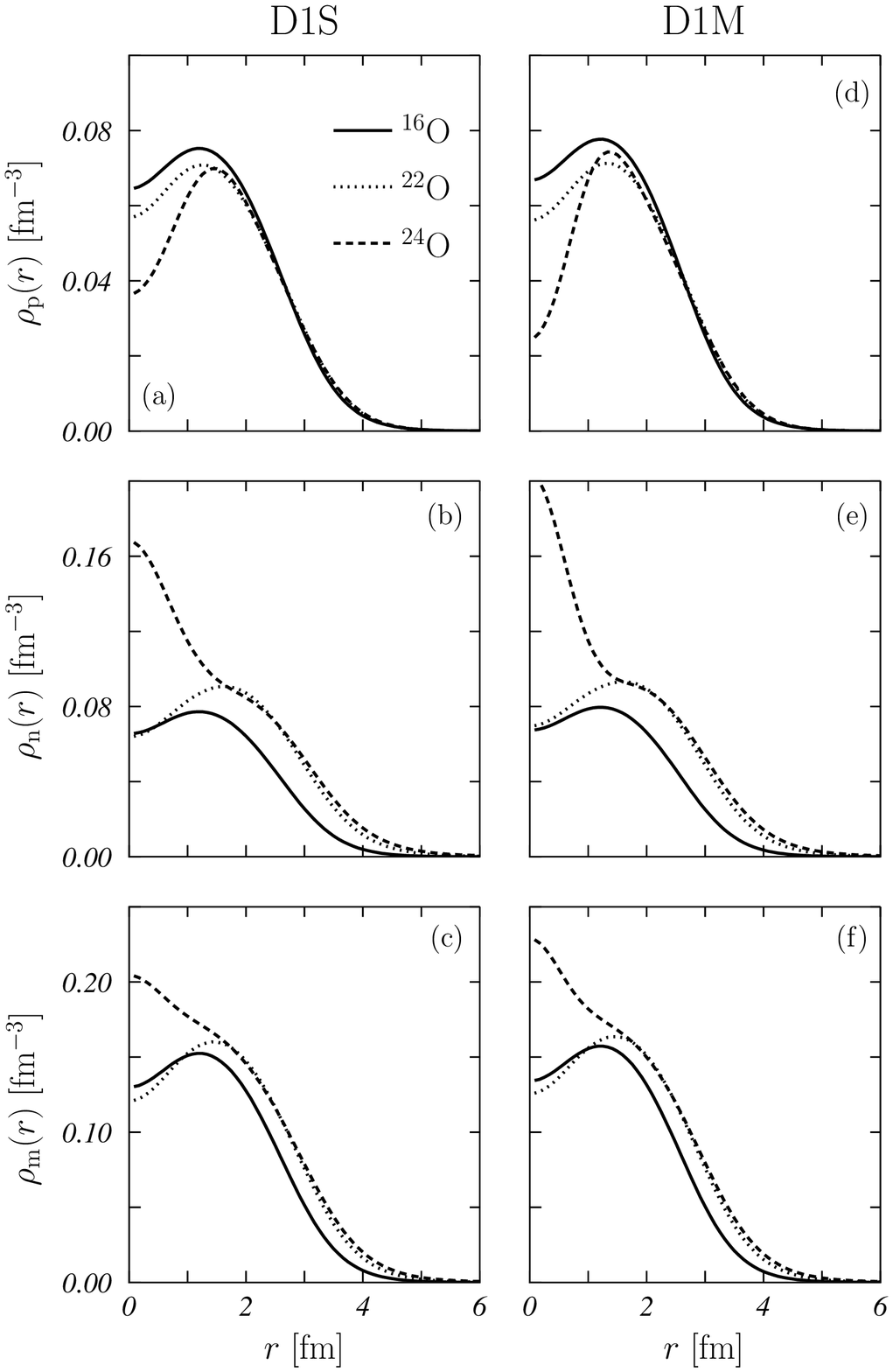} 
\caption{\small Proton $\rho_p$, neutron $\rho_n$ and matter $\rho_m$
  distributions for the three oxygen isotopes we have considered, 
  obtained in HF
  calculations with the D1S and D1M interactions.  }
\label{fig:densox}
\end{center}
\end{figure}

We have studied three oxygen isotopes, the doubly magic \oxy nucleus,
and the $^{22}$O and $^{24}$O isotopes.  In our model, the heavier
isotopes are obtained from the \oxy core by filling, respectively, the
neutron 1d$_{5/2}$, and the 2s$_{1/2}$ s.p. levels. 
The ground state properties of these three isotopes, obtained in HF
calculations with the D1S and D1M interactions, are presented in Table 
\ref{tab:oxene}, and in Figs. \ref{fig:eeox} and  \ref{fig:densox}.

\begin{table}
\begin{center}
\begin{tabular}{lccccccc}
\hline\hline
 & \multicolumn{3}{c}{D1S} &$\quad$  &\multicolumn{3}{c}{D1M} \\ \cline{2-4} \cline{6-8}
      & $^{16}$O &  $^{22}$O &  $^{24}$O &$\quad$  & $^{16}$O & $^{22}$O & $^{24}$O \\
\hline
 $E/A$  & -8.093   &  -7.372 &  -7.012 & $\quad$  & -7.955 & -7.254 & -6.912 \\ \hline
protons    &        &         &   &      &        &        &   \\
$1s_{1/2}$ & -35.37 & -46.43  & -48.64 & $\quad$ & -32.74 & -43.38 & -45.65 \\
$1p_{3/2}$ & -18.58 & -29.89  & -32.32 & $\quad$ & -17.63 & -28.71 & -31.18 \\
$1p_{1/2}$ & -12.49 & -23.99  & -25.97 & $\quad$ & -11.91 & -23.60 & -25.60 \\ \hline
neutrons    &        &        &   &      &        &        &   \\
$1s_{1/2}$ & -38.61 & -41.02  & -41.11 & $\quad$ & -36.00 & -38.32 & -38.46 \\
$1p_{3/2}$ & -21.82 & -22.11  & -22.43 & $\quad$ & -20.91 & -20.72 & -21.15 \\
$1p_{1/2}$ & -15.63 & -18.53  & -17.22 & $\quad$ & -15.10 & -17.35 & -16.29 \\
$1d_{5/2}$ &        &  -6.56  &  -7.01 & $\quad$ &        &  -6.34 & -6.85  \\
$2s_{1/2}$ &        &         &  -4.13 & $\quad$ &        &        & -4.09 \\
\hline\hline
\end{tabular}
\end{center}
\caption{\small Nuclear binding energies per nucleon, $E/A$, 
  and s.p. energies of the
  three oxygen isotopes we have considered, calculated within the HF
  approach by using the D1S and D1M interactions. All the quantities are
  expressed in MeV. The values of the experimental binding energies are
-7.976, -7.365 and -7.016 MeV for $^{16}$O,   $^{22}$O and  $^{24}$O,
  respectively. 
}
\label{tab:oxene}
\end{table}

%
%

%

We show in Table \ref{tab:oxene} the binding energies per nucleon,
$E/A$, and the s.p. energies of the three oxygen isotopes.  The
agreement with the experimental binding energies, whose values are
given in the table caption, is within the few percent.  We observe
that the D1M interaction is slightly less attractive than the D1S
one. In any case, these are relatively small differences and we may
state that, despite the fact that the two interactions are rather
different, as we have shown in Fig. \ref{fig:force}, they give very
similar descriptions of the binding and s.p. energies of the three
nuclei considered. The results of the table show that the proton
s.p. states become more bound with the increasing of the number of
neutrons.  

In the panel (a) of Fig. \ref{fig:eeox} we show the \oxy charge
distributions obtained in HF calculations where the D1S and D1M forces
have been used, and we compare them with the empirical charge
distribution taken from Ref. \cite{dej87}.  Our charge distributions
have been obtained by folding the proton distributions with a dipole
proton electromagnetic form factor. We have verified that more modern,
and accurate, form factors do not produce sizable differences in our
results.  Our calculations describe well the empirical charge
density on the surface but they underestimate it 
in the center of the nucleus.  We show in panel (b) of
Fig. \ref{fig:eeox} the elastic electron scattering cross section
calculated in Distorted Wave Born Approximation \cite{ann95a} by using
the charge distributions shown in the panel (a). We compared our
results with the data of Ref. \cite{sic70,sic}.  The differences
between theoretical and empirical densities show up in the cross
sections at large momentum transfer.  The charge distribution obtained
with the Woods-Saxon potential by using the parameters of
Ref. \cite{ari07}, does not make a better job in describing the
empirical density, which is now overestimated in the nuclear center.

In Fig. \ref{fig:densox}, we show the proton, neutron, and matter 
distributions, $\rho_p$, $\rho_n$ and $\rho_m$ respectively, for the
three oxygen isotopes we have considered. The matter distribution is
the sum of the proton and neutron ones.  We do not remark relevant
differences between the results obtained with the two different
interactions.  The shapes of the neutron distributions show the effects
of the filling of the s.p. levels, which are empty in the lighter
isotopes.  In the $^{22}$O nucleus the $1d_{5/2}$ level, empty in
\oxy, is completely occupied. This level gives a contribution mainly
on the surface. The rms radius of the neutron distribution changes
from 2.64 fm in \oxy to 3.00 fm in $^{22}$O, for the D1S interaction,
and from 2.61 fm to 2.97 fm for the D1M interaction. The situation is
different in $^{24}$O, where the new s.p. level to be occupied is the
$2s_{1/2}$. In this case, the main effect is in the center of the
nucleus. The values of the neutron distributions rms radii are 3.17
fm and 3.12 fm for the D1S and D1M interaction respectively. It is a
relatively small change on the neutron distribution surface.

The proton distributions are interesting since there is no change in
the occupation of the s.p. levels in the different isotopes, therefore
all the differences are produced by the interaction between protons
and neutrons. In the panels (a) and (d) of
Fig. \ref{fig:densox} we show the proton distributions of the three
oxygen isotopes. We notice that the increase of the neutron number
produces a change in the interior of the nucleus. The proton
$1s_{1/2}$ s.p. wave function becomes wider the heavier is the
isotope, and since the normalization is conserved, the value of wave
function at the center of the nucleus becomes smaller. The rms radius
of these distributions changes from 2.20 fm in \oxy to 2.45 fm in
$^{24}$O almost independently of the interaction used. The relevant
lowering of the proton distribution in the nuclear center is produced
to compensate the increase of the neutron density (see panels (b) and
(e)). In the panels (c) and (f) we also show the matter
distributions and we observe that the differences between the various
isotopes are smaller than those shown separately by the proton and
neutron distributions.

So far, we have presented HF results which are related to the ground
states properties of the three oxygen isotopes. We discuss now the
excitation spectra obtained by our CRPA calculations. A first point we
have investigated is related to the relevance of the proper treatment
of the continuum in the self-consistent CRPA calculations. Our study
has been conducted by comparing our CRPA results with the results of
discrete RPA calculations, such as those of Ref. \cite{don09}.  This
is the same strategy adopted in Ref. \cite{nak09} and, within a
relativistic framework, in Ref. \cite{dao09}.  The discrete set of
s.p. states is obtained by solving the HF equations in a box with
bound state boundary conditions. For all oxygen isotopes we use a box
radius of 12 fm. Larger values of this radius do not change binding
and s.p. energies up to the fifth significant figure. There is not
such a stability for the unbound, $\epsilon_p>0$, s.p. wave functions
and energies.  We have controlled the stability of the RPA results by
selecting the maximum value of the particle-hole excitation energy,
$\epsilon^{\rm max}_{ph}$, used in the RPA calculation.  For a given
total angular momentum and parity of the excitation, this value
determines the number of s.p. states forming the configuration space
of the discrete calculation.

The quantity we have considered for these convergence tests is the 
centroid energy, which we calculate as 
\beq
\langle \omega \rangle_J \, = \, 
\displaystyle 
\frac{\displaystyle 
\int_{\omega_{\rm min}}^{\omega_{\rm max}} {\rm d} \omega \, \omega \, 
 B(\omega,EJ:0\to J) }
{ \displaystyle\int_{\omega_{\rm min}}^{\omega_{\rm max}}  
{\rm d}\omega \,  B(\omega,EJ:0\to J) }  
\, .
\label{eq:centr}
\eeq

We have studied the convergence for the $1^{-}$ and $2^{+}$
excitations in all the nuclei we have investigated, and we have found
that the change from $\epsilon^{\rm max}_{ph}$=200 MeV to
$\epsilon^{\rm max}_{ph}$=250 MeV modifies the value of the centroid
energies for less than one part on a thousand. All the discrete RPA
results we present here have been obtained by using $\epsilon^{\rm
max}_{ph}$=250 MeV. Our choice ensures the convergence of discretized
RPA calculations done with HF basis which has been generated by using
a specific value of the box size. 

%
\begin{figure}[ht]
\begin{center}
\includegraphics[scale=0.4,angle=90]{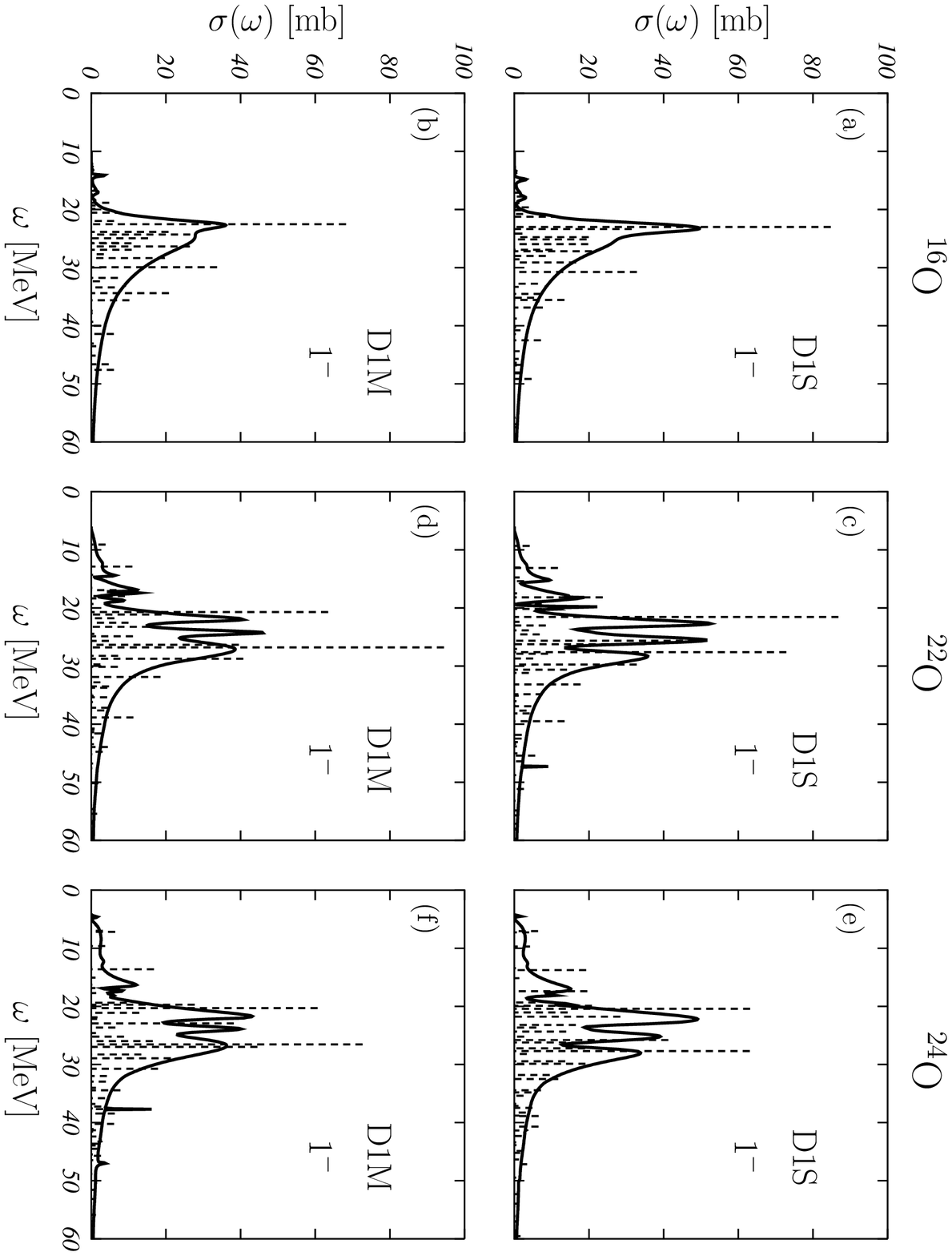} 
\caption{\small Total photoabsorption cross sections calculated 
  with discrete and continuum RPA approaches. 
  The vertical bars show the discrete results, and the solid
  lines those of the CRPA calculations. The excitation multipole 
  is the $1^-$. The upper panels show the results obtained with the
  D1S interaction, the lower panels with the D1M interaction. 
  }
\label{fig:iox1m}
\end{center}
\end{figure}
%

%
\begin{figure}[ht]
\begin{center}
\includegraphics[scale=0.4,angle=90]{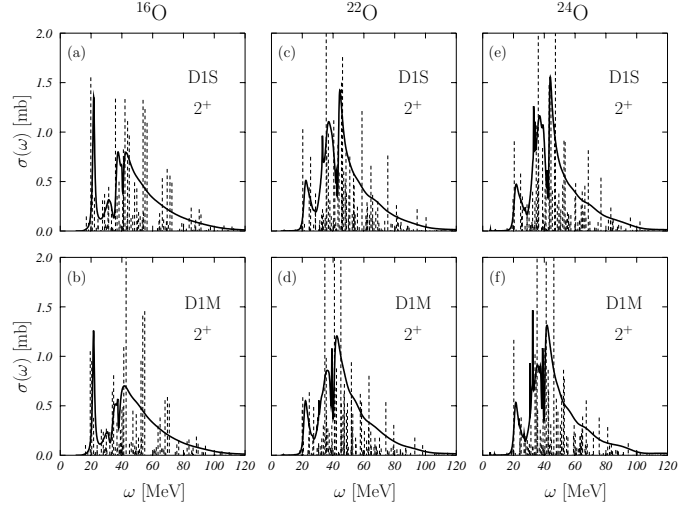} 
\caption{\small The same as Fig. \ref{fig:iox1m} for the $2^+$
  multipole excitation. 
  }
\label{fig:iox2p}
\end{center}
\end{figure}

In Fig. \ref{fig:iox1m} we compare the total photoabsorption cross
sections calculated for the excitation of the $1^{-}$ resonance in the
three oxygen isotopes we have studied.  The vertical bars show the
discrete RPA results and the solid lines those of the CRPA
calculations. In the upper panels we present the results obtained with
the D1S interaction, and in the lower panels those obtained with the
D1M interaction.  

In the \oxy nucleus, the agreement between the results of the two
different calculations is rather good. Discrete results have their
maxima in the same position of those of the continuous
solutions. There are peaks around 30, 35 and 40 MeV which do not have
corresponding partners in the CRPA cross sections. The D1M cross
sections are slightly smaller, indicating, again, that this
interaction is less attractive than the D1S force. The situation is
more complicated in $^{22}$O.  Discrete and continuum results have
similar structures, but the positions of the peaks are slightly
different. In any case, the cross sections show a richer structure
than in the \oxy case. This situation is worsening in $^{24}$O where
the peaks of the continuous cross sections do not correspond to those
of the discrete calculation.

We show in Fig. \ref{fig:iox2p} analogous results for the excitation
of the $2^{+}$ resonance. In this case the results of the discrete RPA
are rather different from those of the CRPA, even in the \oxy nucleus.
The discrete calculations show clusters of peaks not present in the
continuous calculations.  

%
\begin{table}[htb]
\begin{center}
\begin{tabular}{ccccccc}
\hline \hline
 &~~& $^{16}$O &~~& $^{22}$O &~~& $^{24}$O \\ \hline
  1$^-$  &&          &&           &&       \\
RPA-D1S  && 26.34    && 24.87     && 22.82 \\
CRPA-D1S && 27.17    && 25.11     && 23.07 \\
RPA-D1M  && 26.36    && 24.72     && 22.81 \\
CRPA-D1M && 27.23    && 25.04     && 23.20 \\ \hline
  2$^+$  &&          &&           &&       \\
RPA-D1S  && 28.43    && 32.91     && 31.58 \\
CRPA-D1S && 30.47    && 33.47     && 32.20 \\
RPA-D1M  && 28.19    && 32.26     && 30.93 \\
CRPA-D1M && 30.51    && 33.06     && 31.06 \\
\hline \hline
\end{tabular}
\vskip 0.5 cm 
\caption{\small Centroid energies in MeV, Eq. (\ref{eq:centr}),
  for the 1$^-$ and 2$^+$
  electromagnetic excitations in the three oxygen
  isotopes we have studied, obtained with discrete (RPA)
  and continuum (CRPA) calculations. 
}
\label{tab:cen2}
\end{center} 
\end{table}

We show in Table \ref{tab:cen2} the centroid energies of the
electromagnetic responses obtained in discrete and continuum RPA
calculations. We have considered for ${\omega_{\rm min}}$ the values
corresponding to the continuum thresholds. The $1^{-}$ resonances have
been integrated up to ${\omega_{\rm max}}$= 60 MeV, while the $2^{+}$
resonances up to ${\omega_{\rm max}}$= 120 MeV.  The relative
differences between these centroid energies are smaller than 2\% in
the $1^-$ case. In the case of the $2^+$ excitation we reach the
maximum value of 4\% relative difference between the D1M results in
$^{16}$O.

%
\begin{figure}[ht]
\begin{center}
\includegraphics[scale=0.4,angle=90]{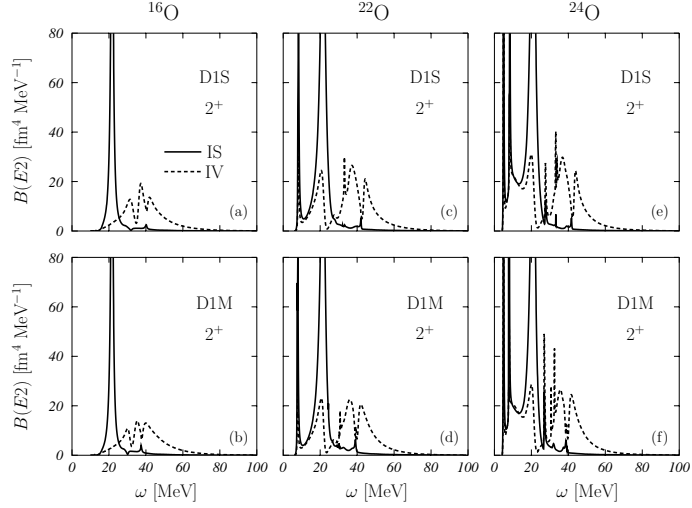} 
\caption{\small 
  Isoscalar (IS) and isovector (IV) strengths of the 2$^+$ excitation
  mode for the oxygen isotopes we have studied. 
  }
\label{fig:isivox}
\end{center}
\end{figure}

While the $1^{-}$ response to photon excitations is essentially of pure 
isovector character, the $2^+$ response is a combination of isoscalar
(IS) and isovector (IV) modes. For the $2^+$ excitation we have
separated the IS and IV responses by including in the expression of
the $B(EJ)$ value of Eq. (\ref{eq:bv}) the operators
\begin{eqnarray}
 T^{IS}_{JM} &=& \sum_{i=1}^A \, r_i^J \, Y_{JM}(\Omega_i) \, \\
 T^{IV}_{JM} &=& \sum_{i=1}^A \, r_i^J \, Y_{JM}(\Omega_i) \, \tau_3(i)
\label{eq:isiv}
\end{eqnarray}
with $J=2$. We show in Fig. \ref{fig:isivox} the IS and IV responses
for the $2^+$ excitation of all the three oxygen isotopes we have
investigated obtained by using the D1S and the D1M interactions. The
IS responses are concentrated at lower energies and show a sharp peak,
while the IV responses are broader and they have less pronounced
maxima at higher energies. In the \oxy nucleus the IS quadrupole
resonance has been identified in $\alpha$ scattering processes at a
peak energy of about 21 MeV \cite{kno75}, to be compared with the peak
energies of 21.7 and 21.6 MeV obtained with the D1S and D1M forces
respectively. The centroid energies of our calculations for the \oxy
nucleus, calculated for $\omega_{\rm max}$=100 MeV are 22.94 and 23.13
MeV for D1S and D1M interactions respectively. If we consider
$\omega_{\rm max}$=40 MeV we obtain 22.12 and 22.24 MeV for these
centroid energies.

%
\begin{figure}[ht]
\begin{center}
\includegraphics[scale=0.4,angle=90]{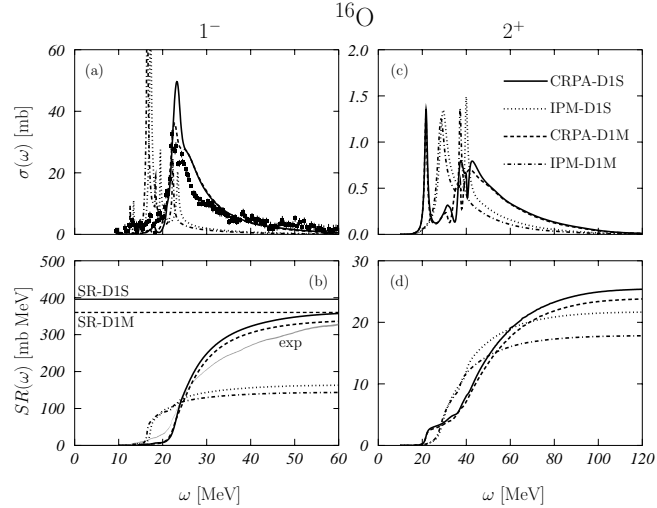} 
\caption{\small Total photoabsorption cross sections for the $1^-$
  (panel (a)) and $2^+$ (panel (c)) excitations. Experimental data
  (solid squares) are from Ref. \cite{ahr75}. In panels
  (b) and (d), we show the sum rule exhaustion functions,
  Eq. (\ref{eq:sr}), for the two multipole excitations. The grey area
  indicates the same function calculated for the experimental
  data. Full and dashed lines show the CRPA results obtained with the
  D1S and D1M interactions, respectively, and the dotted and
  dashed-dotted lines the IPM results obtained with the D1S and D1M HF
  single particle wave functions, respectively.  The horizontal lines
  indicate the sum rule limiting values obtained by using a nuclear
  matter estimate of the enhancement factor. The values of the TRK sum
  rule is 240 mb MeV.  }
\label{fig:mfox}
\end{center}
\end{figure}

Another point we have investigated is related to the effects of the
residual interaction in CRPA calculations. In Fig. \ref{fig:mfox} the
CRPA results (solid and dashed curves) are compared with the IPM
results (dotted and dashed-dotted curves), obtained by switching off
the residual interaction in the CRPA calculation, and with the data of
Ref. \cite{ahr75}.  Since more than the 90\% of the contribution to
the total photoabsorption cross section is given by the $1^-$
excitation, we compare the data with results of this excitation
mode. The contribution of the $2^+$ excitation to the photoabsorption
cross section is shown in the panel (c) of the figure, and it is one
order of magnitude smaller than the contribution of the $1^-$ mode. In
the lower panels we present the sum rule exhaustion functions, 
\beq
SR(\omega)\, = \, \int_0^\omega {\rm d}\omega' \, \sigma(\omega') 
\label{eq:sr}
\, ,
\eeq
calculated for the cross sections shown in the upper panels.

The results obtained with the D1S interaction do not show significant
differences with respect to those obtained with the D1M
interaction. Evidently, only the CRPA calculations predict the
presence, and also the positions, of the resonances. The positions of
the peaks are well reproduced for both multipole excitations by the
CRPA calculations, while the IPM results do not give a good
description of the data.  The sum rule functions obtained with the IPM
calculations are smaller than those of the CRPA. The differences are
larger for the $1^-$ excitation than for the $2^+$ excitation.  This
may be due to the different isospin character of the two excitations,
which, as we have already pointed out, is mainly IV in the $1^-$ mode,
and a combination of IS and IV in the $2^+$ mode.  In the $1^-$ case,
the energies of the peaks of the CRPA cross sections are larger than
those of the IPM results.  This because the residual interactions are
repulsive in the isovector channel. The situation is inverted for the
lower energy peaks of the $2^+$ excitation mode.  The CRPA cross
sections present peaks at lower energies than those of the IPM
calculations.  This indicates that the interactions are attractive in
the isoscalar channel.  The other, wider, $2^+$ resonances peaked at
energies of about 42 MeV, have, instead, IV character, and their
energies are slightly greater than those of the IPM ones.

The comparison of the CRPA results with the photoabsorption data
emphasizes the well known limitations of the RPA description of the
giant resonances. The strength is too concentrated in the peak region,
and the data show a wider energy distribution. The sum rule functions
of the panel (b) of Fig. \ref{fig:mfox} further confirm these
deficiencies.  Even though experimental and CRPA curves seem to have
the same limiting values, the CRPA curves saturate much earlier than
the experimental one.  Again, the strength is too concentrated in the
resonance region.

The saturation value given by the Thomas-Reiche-Khun (TRK) sum rule is
of 240 mb MeV. The isospin dependence of our interactions is
responsible for the fact that our calculations saturate at higher
values. The calculation of the enhancement factor $\kappa$ of the TRK
sum rule is rather involved for finite-range interactions
\cite{tra87}. We use the values of $\kappa$ for the D1S and D1M
interactions obtained by a nuclear matter estimate \cite{nak09}. We
obtain $\kappa$=0.65 for the D1S interaction and  $\kappa$=0.50 for
the D1M force. These values correspond in \oxy to sum rule limiting
values of 396 and 360 mb MeV for the D1S and D1M interaction
respectively. As we show in the panel (b) of Fig. \ref{fig:mfox},
these values are compatible with the results we obtain with our CRPA
calculations.  
 
%
\begin{figure}[ht]
\begin{center}
\includegraphics[scale=0.4,angle=90]{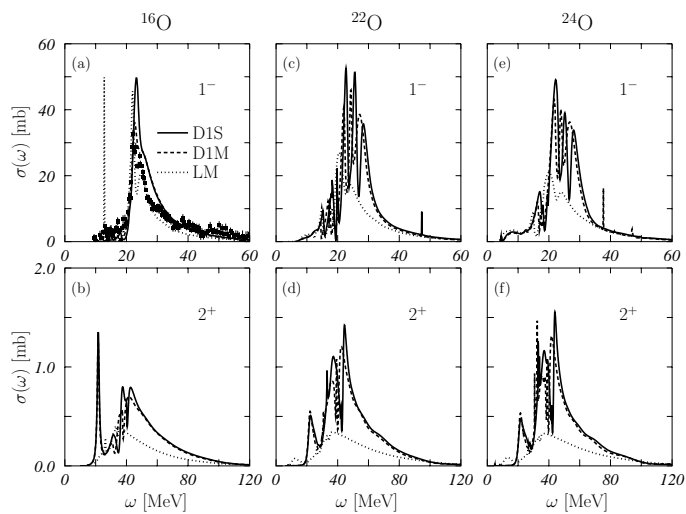} 
\caption{\small Comparison between the self-consistent CRPA results
  obtained with the D1S interaction (full lines), the D1M interaction
  (dashed lines) and those obtained with the phenomenological approach
  of Refs. \cite{deh82,co85}  
  where a Landau-Migdal interaction has been used (dotted lines). 
  }
\label{fig:allox}
\end{center}
\end{figure}

Another issue we have investigated is the capacity of a
phenomenological CRPA approach to predict the excitation spectra of
experimentally unknown nuclei. For this reason we have calculated the
$1^-$ and $2^+$ excitations of the three oxygen isotopes by using the
phenomenological approach of Refs. \cite{deh82,co85}.  In the
phenomenological 
calculations we solved the CRPA equations without exchange terms by
using a zero-range Landau-Migdal force, whose parameters are those of
Ref. \cite{co09b}.  In this approach the s.p. wave functions are
generated by solving the one-body Schr{\"o}dinger equation with a
Woods-Saxon potential.  For all the oxygen isotopes, we used the
parameters of the Woods-Saxon potential of \oxy given in \cite{ari07}.
In Fig. \ref{fig:allox} we compare our self-consistent CRPA results
with those of the phenomenological approach which are indicated by the
dotted lines.

The phenomenological results compare rather well with experimental
data \cite{ahr75} and with our self-consistent CRPA results in the
case of the $1^-$ excitation of the \oxy nucleus.  The position of the
peak coincides with that obtained in the self-consistent calculations,
and all of them are rather close to the experimental one. We remark,
however, that the global strength of the phenomenological result is
smaller than that produced in the self-consistent approach. The
reasonable agreement between the CRPA results obtained in the case of
the $1^-$ excitation in $^{16}$O is peculiar since all the other cases
show large differences between phenomenological and self-consistent
results. The phenomenological calculation predicts the isoscalar $2^+$
excitation of \oxy at 21 MeV energy, but it fails in describing the
isovector excitation at higher energies. The differences between the
results of the phenomenological and self-consistent calculations in
the other two oxygen isotopes are large. The total strengths of the
phenomenological cross sections are much smaller than those of the
self-consistent ones. Great part of the resonance structure of the
self-consistent cross sections is absent in the phenomenological
results.

\subsection{Calcium}
\label{sec:calcium} 
\begin{figure}[ht]
\begin{center}
\includegraphics[scale=0.4,angle=90]{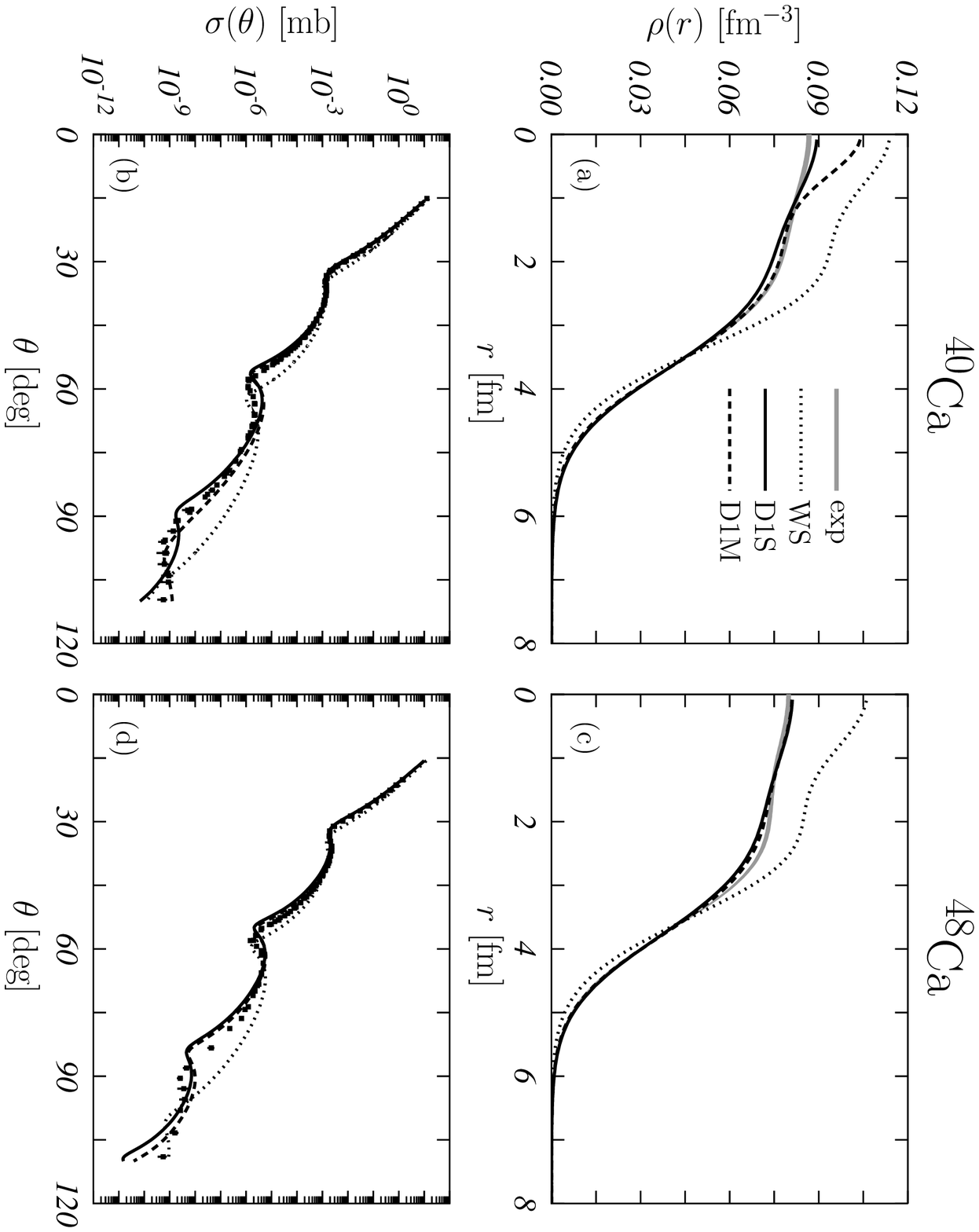} 
\caption{\small The same as Fig. \ref{fig:eeox} for the \caI and \caII
    nuclei. The empirical densities are taken from the compilation of
    Ref. \cite{dej87}. The elastic electron scattering data on \caI 
    from Refs. \cite{sin73,sic79,cav80t}, and those of \caII 
    from Refs. \cite{sic75a,sic}, have been rescaled to match a unique
    electron energy of 400 MeV. 
    }
\label{fig:eeca}
\end{center}
\end{figure}

\begin{figure}[ht]
\begin{center}
\includegraphics[scale=0.4] {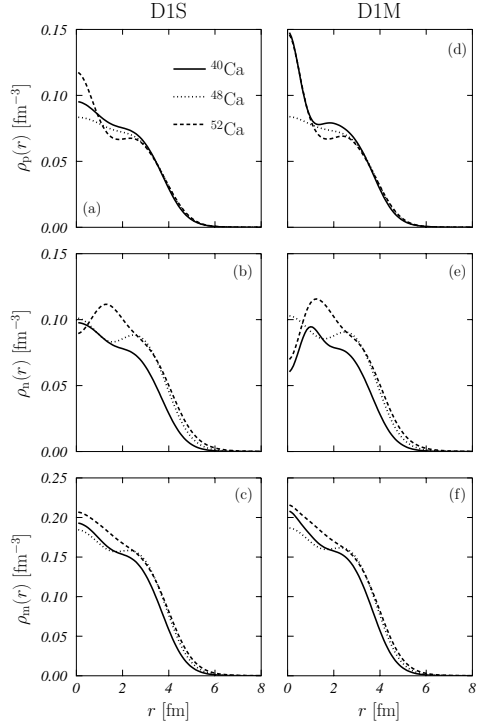} 
\caption{\small Proton $\rho_p$, neutron $\rho_n$ and matter $\rho_m$
  distributions for the three calcium isotopes, obtained in HF
  calculations with the D1S and D1M interactions.
  }
\label{fig:densca}
\end{center}
\end{figure}

The same type of investigation done for the oxygen isotopes has been
repeated for three calcium isotopes: $^{40}$Ca, $^{48}$Ca and
$^{52}$Ca.  The ground state properties of these nuclei are presented
in Table \ref{tab:caene} and in Figs. \ref{fig:eeca} and
\ref{fig:densca}. In Table \ref{tab:caene} we give the values of the
binding and s.p. energies. As in case of oxygen, the agreement with
the experimental binding energies is within few percents. To be
precise, we remark that the D1S results are slightly better than those
obtained with the D1M force. Also in these calculations the D1M force
shows less attraction than the D1S interaction. As in the case of
oxygen, the proton s.p. levels become more bound with increasing
number of neutrons.

\begin{table}
\begin{center}
\begin{tabular}{lccccccc}
\hline\hline
 & \multicolumn{3}{c}{D1S} &$\quad$ &  \multicolumn{3}{c}{D1M} \\ \cline{2-4} \cline{6-8}
 & $^{40}$Ca &  $^{48}$Ca &  $^{52}$Ca &$\quad$ & $^{40}$Ca & $^{48}$Ca & $^{52}$Ca \\ \hline
 $E/A$       & -8.579 & -8.639  & -8.344 &$\quad$ & -8.462 & -8.537 & -8.260 \\ \hline
protons    &        &         &        &$\quad$ &        &        &   \\
$1s_{1/2}$ & -44.82 & -51.31  & -54.01 &$\quad$ & -41.01 & -46.90 & -49.35 \\
$1p_{3/2}$ & -30.04 & -37.61  & -40.02 &$\quad$ & -27.84 & -34.91 & -37.16 \\
$1p_{1/2}$ & -26.05 & -33.89  & -35.76 &$\quad$ & -24.14 & -31.74 & -33.60 \\
$1d_{5/2}$ & -16.02 & -23.89  & -26.37 &$\quad$ & -15.15 & -22.65 & -25.07 \\
$2s_{1/2}$ & -10.52 & -17.12  & -20.93 &$\quad$ &  -9.91 & -16.17 & -19.98 \\
$1d_{3/2}$ &  -9.18 & -16.96  & -19.29 &$\quad$ &  -8.75 & -16.54 & -18.82 \\ \hline
neutrons   &        &         &        &$\quad$ &        &        &   \\
$1s_{1/2}$ & -52.07 & -53.20  & -53.14 &$\quad$ & -48.44 & -49.57 & -49.52 \\
$1p_{3/2}$ & -37.09 & -37.89  & -37.88 &$\quad$ & -35.00 & -35.60 & -35.71 \\
$1p_{1/2}$ & -33.01 & -35.32  & -34.46 &$\quad$ & -31.21 & -33.23 & -32.56 \\
$1d_{5/2}$ & -22.96 & -22.92  & -23.09 &$\quad$ & -22.15 & -21.84 & -22.13   \\
$2s_{1/2}$ & -17.54 & -18.27  & -17.40 &$\quad$ & -17.03 & -17.69 & -17.89 \\
$1d_{3/2}$ & -15.95 & -17.83  & -18.85 &$\quad$ & -15.57 & -17.09 & -16.83 \\
$1f_{7/2}$ &        &  -9.39  &  -9.76 &$\quad$ &        &  -9.25 &  -9.71 \\
$2p_{3/2}$ &        &         &  -5.49 &$\quad$ &        &        &  -5.50 \\
\hline\hline
\end{tabular}
\end{center}
\caption{\small Nuclear binding energies per nucleon, $E/A$, 
  and s.p. energies of the
  three calcium isotopes we have considered, calculated within the HF
  approach by using the D1S and D1M interactions. All the quantities are
  expressed in MeV. The values of the 
  experimental binding energies are
  -8.551, -8.666 and -8.396 MeV for $^{40}$Ca,   $^{48}$Ca and  $^{52}$Ca,
  respectively. 
}
\label{tab:caene}
\end{table}

%

In the upper panels of Fig. \ref{fig:eeca} we show the charge
distributions of \caI and \caII nuclei. We use these charge
distributions to calculate in Distorted Wave Born Approximation the
elastic electron scattering cross sections. These cross sections are
compared with the experimental data
\cite{sin73,sic79,cav80t,sic75a,sic} in the lower panels of the
figure. In the figure we also show, with the dotted lines, the charge
densities and the associated cross sections obtained from
phenomenological calculations done by using the Woods-Saxon potential
with the parameters given in Ref. \cite{ari07}.  The dashed-dotted
lines showing the empirical densities are taken from the compilation
of Ref.  \cite{dej87}.  The results of our HF calculations show a
better agreement with the data than those of the phenomenological
calculations.

%

The proton, neutron and matter distributions of the three calcium
isotopes are shown in Fig. \ref{fig:densca}. The main features pointed
out in the discussion done for the oxygen isotopes are present also in
this case, where there are, however, remarkable differences in the
details. In the oxygen isotopes the $2s_{1/2}$ state was occupied only
by the neutrons in the $^{24}$O nucleus. In the calcium isotopes we
have considered, the $2s_{1/2}$ state is always occupied in both
proton and neutron cases.  The \caII is obtained from the \caI nucleus
by filling the neutron $1f_{7/2}$ state, and the $^{52}$Ca by filling,
in addition, the neutron $2p_{3/2}$ state. In the panel (b) of
Fig. \ref{fig:densca} it is shown that the filling of the $1f_{7/2}$
state increases the neutron surface, and leaves practically unmodified
the density at the center of the nucleus. The filling of the
$2p_{3/2}$ state modifies the neutron density mainly around 2.0 fm,
but it produces also a small lowering at the nuclear center. These
modifications change the proton densities at the nuclear center as it
is shown in the panel (a). The matter distributions obtained with
the D1S interaction, and shown in panel (c), are rather smooth in
the nuclear interior.  The D1M interaction generates in \caI narrow
s-waves, and this produces a large proton distribution in the nuclear
center. The corresponding neutron distribution, panel (e), has a
hole in the center, and this compensates the peak of the proton
distribution and produces a matter distribution rather smooth. The
same type of considerations can be done also for the distributions of
the $^{52}$Ca isotope. In general, we observe that, as in the oxygen
case, the HF calculations find the optimal matter distributions which
is rather smooth, even though the separated proton and neutron
densities may show some rapid changes.

%
\begin{figure}[ht]
\begin{center}
\includegraphics[scale=0.4,angle=90]{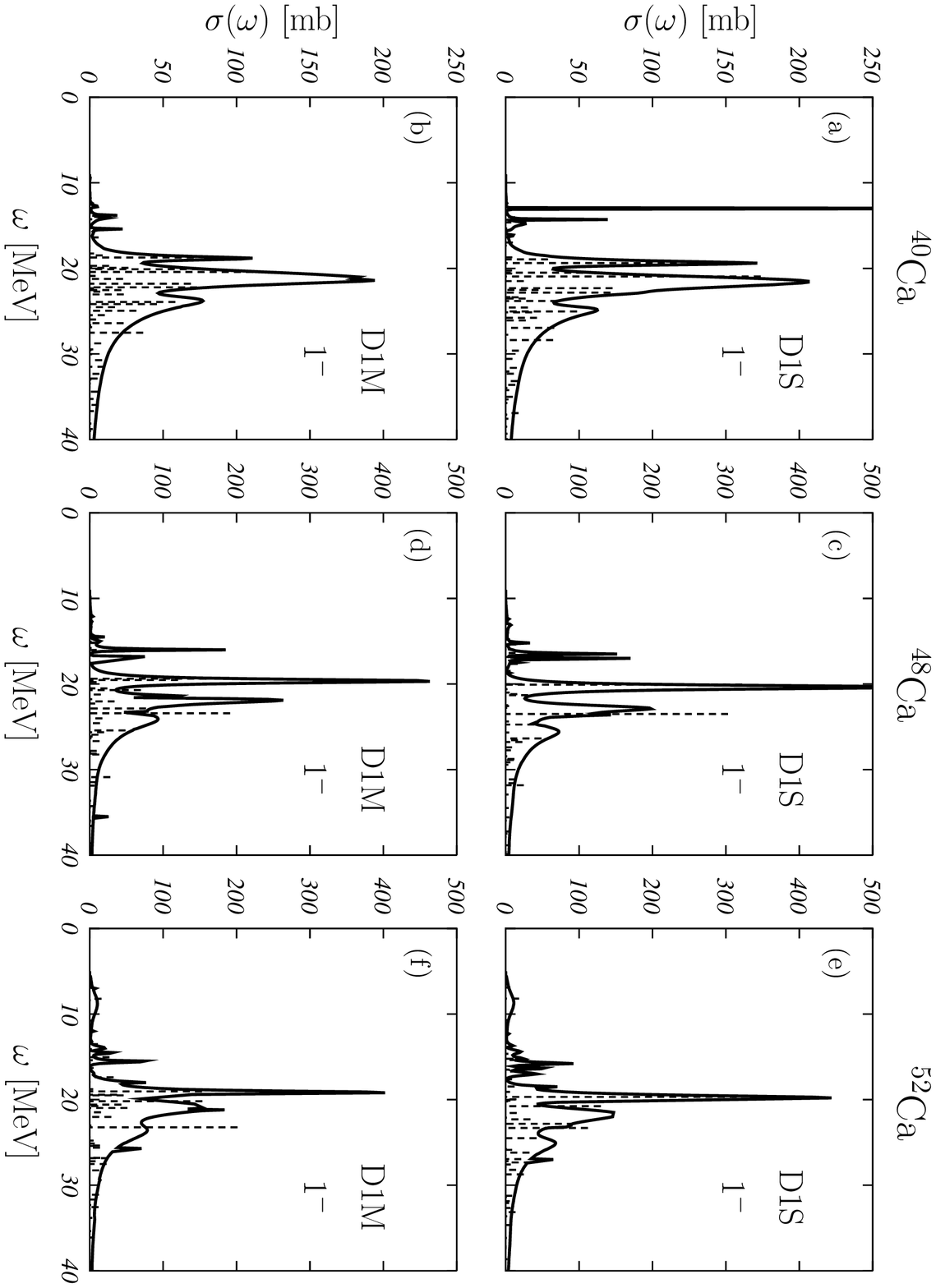} 
\caption{\small The same as Fig. \ref{fig:iox1m} for the three calcium
  isotopes. Note that the vertical scale for \caI is half that of
  \caII and of $^{52}$Ca.
  }
\label{fig:ica1m}
\end{center}
\end{figure}
%

%
\begin{figure}[ht]
\begin{center}
\includegraphics[scale=0.4,angle=90]{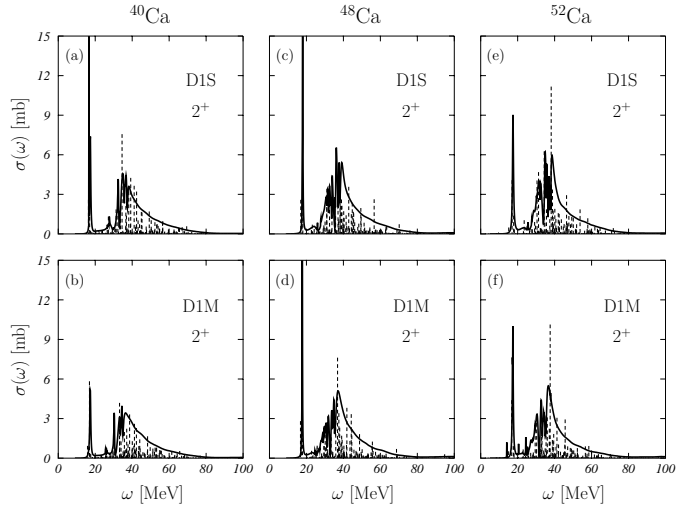} 
\caption{\small The same as Fig. \ref{fig:iox2p} for the three calcium
  isotopes. 
  }
\label{fig:ica2p}
\end{center}
\end{figure}

The comparison between discrete and continuum RPA results is done in
Figs. \ref{fig:ica1m} and \ref{fig:ica2p} for the 1$^-$ and 2$^+$
multipole excitations, respectively.  As in the oxygen case, we show
the contributions to the total photoabsorption cross sections.  The
agreement between the results of discrete and continuum RPA
calculations is slightly worse than in the oxygen case. Both type of
calculations produce resonances, however the total strengths predicted
by the discrete RPA are about one half those of the CRPA. For the
calcium isotopes, this result is common to all the nuclei,
interactions and multipole excitations we have investigated.

%
\begin{table}[htb]
\begin{center}
\begin{tabular}{ccccccc}
\hline \hline
        &~~&  {$^{40}$Ca} &~~& {$^{48}$Ca} &~~& {$^{52}$Ca} \\\hline
  1$^-$ &&              &&             &&   \\
RPA-D1S  && 22.37 && 22.27 && 20.56  \\ 
CRPA-D1S && 21.89 && 22.40 && 20.63  \\
RPA-D1M  && 22.10 && 21.86 && 20.24  \\
CRPA-D1M && 22.42 && 21.77 && 20.33  \\\hline
  2$^+$  &&       &&       &&  \\ 
RPA-D1S  && 24.66 && 27.34 && 26.84 \\
CRPA-D1S && 25.02 && 26.89 && 26.14 \\
RPA-D1M  && 24.39 && 26.70 && 26.12 \\
CRPA-D1M && 27.36 && 26.29 && 25.49 \\
\hline\hline
\end{tabular}
\vskip 0.5 cm 
\caption{\small Centroid energies in MeV, Eq. (\ref{eq:centr}),
  for the 1$^-$ and 2$^+$ electromagnetic excitations
  in the various calcium isotopes. The discrete RPA results have 
  been obtained by using $\epsilon^{\rm max}_{ph}$=250 MeV. 
  }
\label{tab:cenca}
\end{center} 
\end{table}

We show in Table \ref{tab:cenca} the centroid energies of the
electromagnetic responses obtained in discrete and continuum RPA
calculations. As we have done for the oxygen isotopes we have taken
for ${\omega_{\rm min}}$ the values corresponding to the continuum
thresholds for the various calcium isotopes.  The $1^{-}$ resonances
have been integrated up to ${\omega_{\rm max}}$= 40 MeV, while the
$2^{+}$ resonances up to ${\omega_{\rm max}}$= 100 MeV.  The relative
differences between these centroid energies are smaller than 1\% for
the $1^-$ excitations and reach the value of about the 6\% for the
$2^+$ states.  We have studied separately the centroid energies of the
IS and IV components of the $2^+$ excitations. By selecting the same
values of ${\omega_{\rm min}}$ and ${\omega_{\rm max}}$ we reproduce
the energy differences of Ref. \cite{nak09}.

%
\begin{figure}[ht]
\begin{center}
\includegraphics[scale=0.4,angle=90]{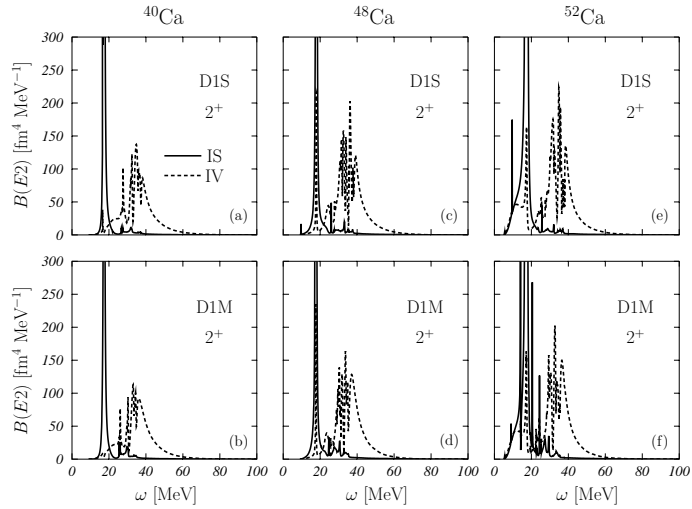} 
\caption{\small 
  IS and IV strengths of the 2$^+$ excitation
  mode for the calcium isotopes we have studied. 
  }
\label{fig:isivca}
\end{center}
\end{figure}

The structure of the electromagnetic 2$^+$ strength distributions is
analogous to that of the oxygen isotopes. They are characterized by a
narrow IS peak at lower energies, here around 17-18 MeV, and a much wider
IV resonance at higher energies. The IS and IV 2$^+$ responses are
separately shown in Fig. \ref{fig:isivca} for the three calcium
isotopes we are studying, and for the two interactions we are using.
The peak of the IS 2$^+$ resonance has been identified in \caI at
17.7$\pm$0.2 MeV in $\alpha$ scattering processes \cite{lui81}. This
value should be compared with the peak energy of 17.4 MeV in CRPA
calculations done with both D1S and D1M forces. The presence of an IV
2$^+$ resonance in \caI around 32 MeV has been indicated in the
analysis of radiative neutron capture data \cite{ber84}. Our
calculations produce wide, and fragmented, IV resonances. We have
peaks around 32 MeV but also around 35 MeV.

%
\begin{figure}[ht]
\begin{center}
\includegraphics[scale=0.4,angle=90]{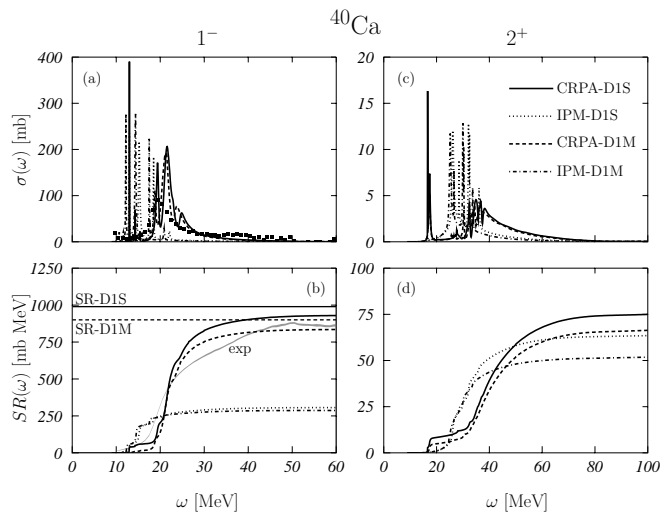} 
\caption{\small The same as Fig. \ref{fig:mfox} for \caI. Experimental data
  (solid squares) are from Ref. \cite{ahr75}. 
  The value of the TRK sum rule is 600 mb MeV. 
  }
\label{fig:mfca}
\end{center}
\end{figure}

In Fig. \ref{fig:mfca} we compare our \caI CRPA results with the total
photoabsorption data of Ref. \cite{ahr75} and with the result of the
IPM calculations. The same observations done for the oxygen case are
valid also here. The IPM results are unable to describe the
experimental cross section.  The sum rule functions (\ref{eq:sr})
shown in the panels (b) and (d) of the figure confirm what we have
observed in the oxygen case.  The strengths of the IPM results are
much smaller than those of the CRPA calculations. In the 1$^-$ case,
the CRPA sum rule function reaches the empirical value but too early
with respect to the empirical behaviour. This indicates that the
strength is too concentrated in the resonance region.

The value of the traditional TRK sum rule is 600 mb MeV, much lower
than the limiting values obtained by our CRPA calculations. The
nuclear matter estimates of the enhancement factors gives limiting
values of the sum rules of 990 and 900 mb MeV for the D1S and D1M
interactions, respectively. We show in the panel (b) of
Fig. \ref{fig:mfca} that these values are compatible with out CRPA
results. 

Also in the case of the calcium isotopes we have compared our
self-consistent CRPA results, with those of the phenomenological
approach. We obtain results analogous to those shown in the oxygen
case. The phenomenological results show less strength and structure
than the self-consistent ones.

\section{Conclusions}   
\label{sec:conclusions}
In this article we presented a technique to solve the CRPA equations
in a self-consistent framework.  In our calculations we used
Gogny-like finite-range interactions containing zero-range density
dependent terms.  Also the spin-orbit term, used only in the HF
calculations, is of zero-range type.  We have shown results for the
$A=16$, 22 and 24 oxygen isotopes, and for $A=40$, 48 and 52 calcium
isotopes. We have compared our results with the available experimental
total photoabsorption data.  We have studied the need of a proper
treatment of the continuum, by comparing our results with those
obtained by discrete RPA calculations. The need of a self-consistent
approach has been investigated by making a comparison with the results
of phenomenological CRPA calculations.

We summarize here below the main results of our study.
\begin{itemize}
\item [-] The D1S and D1M forces are very different if compared in the
  various interaction channels Eq. (\ref{eq:fcahnnels}).  However, they
  produce very similar results, both in HF and in CRPA calculations.
\item [-] In the HF calculations the global matter distribution, given
  by the sum of the proton and neutron distributions, is modified to
  obtain the minimization of the total energy of the system.  We have
  shown in Figs. \ref{fig:densox} and \ref{fig:densca} that the HF
  minimization procedure produces rather smooth matter distributions,
  even though separately, the proton and neutron densities may
  strongly variate.
\item [-] With the increasing number of neutrons, the protons s.p. states
  become more bound, as we have shown in Tabs. \ref{tab:oxene} and
  \ref{tab:caene}. This effect is not relevant for the charge
  conserving excitations, those treated in this work, but it may have
  consequences in charge exchange excitations.
\item [-] The IPM calculations of the nuclear responses are unable to
  provide the proper strength of the multipole excitation. This
  confirms a well known fact that IPM calculations do not predict the
  presence of giant resonances in the nuclear excitation spectrum. 
\item [-] The comparison with discrete RPA results shows the need of a
  correct treatment of the continuum in self-consistent
  calculations. In discrete calculations we have used
  s.p. configuration spaces large enough to ensure the stability of
  the results in the low-lying states and in the giant resonance
  excitation regions.  Discrete RPA calculations can reproduce some
  bulk properties of the excitation, the centroid energies for
  example, but they fails in the detailed description.  Discrete RPA
  responses show clustering of excited stated that the CRPA strength
  distributions do not have.
\item [-] The phenomenological CRPA results are similar to
  those of the self-consistent CRPA calculations in \oxy and \caI. The
  input parameters of the phenomenological calculations have been
  chosen to reproduce some experimental quantities in these nuclei. On
  the contrary, when we apply this approach to the other isotopes, we
  found results which are rather different from those obtained with
  the self-consistent approach.  This indicates the inadequacy of the
  phenomenological approach in the study of nuclei lying in
  experimentally unexplored parts of the nuclear isotope chart.
\item [-] Self-consistent CRPA calculations describe rather well the
  experimental positions of the giant resonance peaks, for both the
  $1^-$ and the $2^+$ excitations.  On the other hand, the strength 
  distributions are incorrect since they are concentrated in the peak
  region, while the experimental distributions are wider. This is a
  well known deficiency of the RPA description of nuclear giant
  resonances. There are strong indications that the problem could be
  solved by considering the excitation of two particle-two hole
  pairs \cite{dro90,kam04,gam10}.  
\end{itemize}

The work presented here is the first step of a project aiming to apply
a self-consistent computational scheme to many other observables and
nuclei. The next step of our work will be the study of unnatural
parity excitations, and for this investigation we shall consider a
tensor term in the interaction \cite{co10}.  It has been shown that
this term slightly affects the ground state properties,
\cite{ots06,mor10} but it has more relevant effects on the spectrum of
magnetic states \cite{co90,don09}. The study of charge-exchange
excitations will be a following step.
 
\appendix
\section{Expansion of the CRPA equations on a basis of
sturmian functions}
\label{app:sexp}
In this appendix we derive Eqs. (\ref{eq:st1}) and
(\ref{eq:st2}) by inserting the expansions 
(\ref{eq:expf})  (\ref{eq:expg})
of the $f$ and $g$
functions on the sturmian functions basis in the CRPA secular equations
(\ref{eq:feq}) and (\ref{eq:geq}).  We insert the expression
(\ref{eq:expf}) in the first term of Eq. (\ref{eq:feq}) and, by using
the definition (\ref{eq:orthost}) of the orthogonalized sturmian
functions and the fact that $R_{p}(r,\epsilon_p)$ is an eigenfunction
of the s.p. hamiltonian $\cah$ (see Eq. (\ref{eq:hf1})) for the
eigenvalue $\epsilon_p = \epsilon_h + \omega$, we obtain
\beqn 
\cah [f^{p_0h_0}_{ph}(r)] \, - \, 
(\epsilon_h \, + \, \omega) \, f^{p_0h_0}_{ph}(r) & = & \cah
\Big[R_{p_0}(r,\epsilon_p)\, \delta_{pp_0}\,
\delta_{hh_0} \, + \, \sum_\mu \, c^{\mu+}_{ph} \,
\widetilde{\Phi}^{\mu+}_{p}(r) \Big] \nonumber \\
 && -(\epsilon_h \, + \, \omega) \, 
\Big[R_{p_0}(r,\epsilon_p)\, \delta_{pp_0}\,
\delta_{hh_0} \, + \, \sum_\mu \, c^{\mu+}_{ph} \,
\widetilde{\Phi}^{\mu+}_{p}(r) \Big] \nonumber \\
& & \hspace*{-2.5cm} = \, \sum_\mu \, c^{\mu+}_{ph} \, 
\Bigg\{ \cah [\Phi^{\mu+}_{p}(r)]
  \, - \, (\epsilon_h \, + \, \omega) \, \Phi^{\mu+}_{p}(r) \\
  \nonumber &~& \hspace*{-0.5cm} - \, \sum_{\epsilon_i<\epsilon_{\rm
      F}} \, \delta_{ip} \, (\epsilon_i \, - \, \epsilon_h\, -\,
  \omega) \, R_i(r) \int {\rm d}r' \, r'^2 \, R_i(r')\,
  \Phi^{\mu+}_p(r') \Bigg\} \, .  \eeqn
The sum of the last term is limited to
the states below the Fermi surface having the same orbital and total 
angular momentum of the particle state. 
Using the definition of the Sturm-Bessel functions given in
Eq. (\ref{eq:sturmb}), we obtain for the above expression
\beqn 
\cah [f^{p_0h_0}_{ph}(r)] \, - \, 
(\epsilon_h \, + \, \omega) \, f^{p_0h_0}_{ph}(r)
&=& \nonumber \\ &~& \hspace*{-4.5cm} =\,-\, \sum_\mu \, c^{\mu+}_{ph} \,
\Bigg\{ \left[ \overline U^\mu_p(r) \, - \, \cau(r) \right] \,
\Phi^{\mu+}_{p}(r) \, + \, \int \, {\rm d}r' \, r'^2 \, \caw(r,r') \,
\Phi^{\mu+}_p (r') \nonumber \\ &~& \hspace*{-2.5cm} + \,
\sum_{\epsilon_i<\epsilon_{\rm F}} \, \delta_{ip} \, (\epsilon_i \, - \,
\epsilon_h\, -\, \omega) \, R_i(r) \int {\rm d}r' \, r'^2 \, R_i(r')\,
\Phi^{\mu+}_p(r') \Bigg\}\, , 
\eeqn 
where we have used the fact that, from Eqs. (\ref{eq:hf1}) and
(\ref{eq:sturmb}), we have
\beq 
\cah \left[\Phi^{\mu+}_p (r) \right] \, - \,
\left[\cau(r) \, + \, \epsilon_p \right] \,
\Phi^{\mu+}_p (r) \, + \, \int \, {\rm d}r' \, r'^2 \, \caw(r,r') \,
\Phi^{\mu+}_p (r')\, \, = \, - \, \overline{U}^\mu_p(r) \,
\Phi^{\mu+}_p(r) \, .  
\eeq
Multiplying the above expression by $r^2\, \Phi^{\nu +}_{p}(r)$
and integrating on $r$ we obtain:
\beqn
\nonumber
\int\,{\rm d}r\, r^2 \, \Phi^{\nu +}_{p}(r) \, \left\{
\cah [f^{p_0h_0}_{ph}(r)] \,
- \, (\epsilon_h \, + \, \omega) \, f^{p_0h_0}_{ph}(r) \right\} \, 
=  &~& \\ &~&
\hspace*{-9.cm} =\, -\, \sum_\mu \, c^{\mu+}_{ph} \, \Bigg\{ \delta_{\mu \nu} \, -
\, \int {\rm d}r \, r^2 \, \Phi^{\nu +}_p(r) \, \Bigg[ \cau(r) \,
\Phi^{\mu+}_p(r) \, - \, \int \, {\rm d}r' \, r'^2 \, \caw(r,r')\,
\Phi^{\mu+}_p (r') \nonumber \\ &~& 
\hspace*{-4.cm} - \, \sum_{\epsilon_i<\epsilon_{\rm F}}
\delta_{ip} \, (\epsilon_i-\epsilon_h-\omega) \, R_i(r)\,  \int
{\rm d}r'\, r'^2 \, R_i(r') \, \Phi^{\mu+}_p(r') \Bigg] \Bigg\} \nonumber \\
&& \hspace*{-9.cm} \equiv \, -\, \sum_\mu \, c^{\mu+}_{ph} \,
\Bigg\{ \delta_{\mu \nu} \, -
\, \langle (\Phi_p^{\nu +})^*|\cau|\Phi_p^{\mu+} \rangle \,+ \, 
\langle (\Phi_p^{\nu +})^*\, \ide |\caw| \ide \,\Phi^{\mu+}_p \rangle \nonumber \\
&& \hspace*{-5.5cm} + \, \sum_{\epsilon_i<\epsilon_{\rm F}}\, \delta_{ip}\, (\epsilon_i-\epsilon_h-\omega)\,
\langle (\Phi_p^{\nu +})^* |R_i\rangle \langle
(R_i)^*|\Phi_p^{\mu+}\rangle \Bigg\}
\label{eq:eqstr1}
\, ,
\eeqn
where we have used the orthogonality relation (\ref{eq:ort1}). The
number of the radial integrations is given by the number of the
functions indicated in the bra and ket symbols. For this reason, in
the terms with $\caw$ we have inserted $\ide$ to indicate the identity
function.

For the right-hand side of Eq. (\ref{eq:feq}), using the orthogonality
relation (\ref{eq:orthost}) we obtain
\beqn 
\int \, {\rm d}r\, r^2 \, \Phi^{\nu +}_{p}(r) \,
\Bigg[
-\, \caf^J_{ph}(r) \, +
\, \sum_{\epsilon_i<\epsilon_{\rm F}} \, \delta_{ip} \, R_i(r) \, 
\int {\rm d}r' \, r'^2 \, R^*_i(r') \, \caf^J_{ph}(r') \Bigg]
\, = &~& \nonumber \\ && \nonumber
\hspace*{-10cm} 
= \, -\, \int \, {\rm d}r\, r^2 \, \Bigg[ \widetilde{\Phi}^{\nu +}_{p}(r) \, + \,
\sum_{\epsilon_i<\epsilon_{\rm F}} \, \delta_{ip} \,
R^*_i(r) \, \int {\rm d}r' \, r'^2 \, R_i(r') \, \Phi^\nu_p(r') \Bigg] \, 
\caf^J_{ph}(r)
\\ && \nonumber 
\hspace*{-9.5cm} 
+ \, \int \, {\rm d}r\, r^2 \, \Phi^{\nu +}_{p}(r) \,
\sum_{\epsilon_i<\epsilon_{\rm F}} \, \delta_{ip} \, R_i(r) \, 
\int {\rm d}r' \, r'^2 \, R^*_i(r') \, \caf^J_{ph}(r') \\
&& 
\hspace*{-10cm} 
= \, -\, \int \, {\rm d}r\, r^2 \, 
\widetilde{\Phi}^{\nu +}_{p}(r) \, \caf^J_{ph}(r)
\,\,.
\eeqn
Using now Eqs. (\ref{eq:Fcal}), (\ref{eq:expf}) and (\ref{eq:expg}) we have
\beqn
-\, \int \, {\rm d}r\, r^2 \, 
\widetilde{\Phi}^{\nu +}_{p}(r) \, \caf^J_{ph}(r) \, = \, 
-\, \int \, {\rm d}r\, r^2 \, \widetilde{\Phi}^{\nu +}_{p}(r) \,
\sum_{p'h'} \, \int {\rm d}r' \, r'^2 && \nonumber \\  
&& \nonumber \hspace*{-8.cm}
\Bigg\{ R^*_{h'}(r') \, \Bigg[
V^{J,{\rm dir}}_{ph,p'h'}(r,r')\, R_h(r) \, f^\nu_{p'h'}(r')
- V^{J,{\rm exc}}_{ph,p'h'}(r,r')\, f^\nu_{p'h'}(r) \, R_h (r') \Bigg]
\\
&&  \hspace*{-8.cm} + \, g^*_{p'h'}(r') \, 
\Bigg[ U^{J,{\rm dir}}_{ph,p'h'}(r,r') \, R_h(r)\, R_{h'}(r')\, -  
 \, U^{J,{\rm exc}}_{ph,p'h'}(r,r') \, R_{h'}(r) \, R_h(r') \Bigg]  
\Bigg\} \nonumber \\ 
&& \nonumber \hspace*{-10.5cm} = \, -\, 
\int \, {\rm d}r\, r^2 \, \widetilde{\Phi}^{\nu +}_{p}(r) \,
\sum_{p'h'} 
\, \int {\rm d}r' \, r'^2  \\ 
&& \nonumber \hspace*{-9.5cm}
\Bigg\{ R^*_{h'}(r') \, V^{J,{\rm dir}}_{ph,p'h'}(r,r')\, R_h(r) \, 
\Bigg[ R_{p_0}(r',\epsilon_{p_0})\, \delta_{p',p_0}\, \delta_{h',h_0} \,  
+\, \sum_\mu \, c^{\mu+}_{p'h'} \, \widetilde{\Phi}^{\mu+}_{p'}(r') \Bigg]
\nonumber \\ 
&& \nonumber \hspace*{-9.2cm}
-\, R^*_{h'}(r') \, V^{J,{\rm exc}}_{ph,p'h'}(r,r')\, \Bigg[
R_{p_0}(r,\epsilon_{p_0})\, \delta_{p',p_0}\, \delta_{h',h_0} \ + \, 
\sum_\mu \, c^{\mu+}_{p'h'} \, \widetilde{\Phi}^{\mu+}_{p'}(r) \Bigg] 
\, R_h(r') \nonumber \\ 
&& \nonumber \hspace*{-9.2cm} 
+ \, \sum_\mu \, (c^{\mu-}_{p'h'})^* \, (\widetilde{\Phi}^{\mu-}_{p'}(r'))^* 
\\  && \nonumber \hspace*{-8.2cm}  \Bigg[ U^{J,{\rm dir}}_{ph,p'h'}(r,r') \, R_h(r)\, R_{h'}(r')\, -  
 \, U^{J,{\rm exc}}_{ph,p'h'}(r,r') \, R_{h'}(r) \, R_h(r') \Bigg]  
\Bigg\}\\ 
&& \nonumber \hspace*{-10.5cm}
= \, -\, \langle (\widetilde{\Phi}_p^{\nu +})^* R_{h_0} |
   V^{J,{\rm dir}}_{ph,p_0h_0} |R_h R_{p_0}(\epsilon_{p_0}) \rangle \,
- \, \sum_{p'h'} \, \sum_\mu \, c^{\mu+}_{p'h'} 
\langle (\widetilde{\Phi}_p^{\nu +})^* R_{h'} |  
V^{J,{\rm dir}}_{ph,p'h'} |R_h \widetilde{\Phi}^{\mu+}_{p'} \rangle \\ 
&& \nonumber \hspace*{-10.cm}
+ \, \langle (\widetilde{\Phi}_p^{\nu +})^* R_{h_0} |
   V^{J,{\rm exc}}_{ph,p_0h_0} |R_{p_0}(\epsilon_{p_0}) R_h  \rangle \, 
+ \, \sum_{p'h'} \, \sum_\mu \, c^{\mu+}_{p'h'} 
\langle (\widetilde{\Phi}_p^{\nu +})^* R_{h'} |  
V^{J,{\rm exc}}_{ph,p'h'} |\widetilde{\Phi}^{\mu+}_{p'} R_h  \rangle \\ 
&&  \hspace*{-10.cm}
- \, \sum_{p'h'} \, \sum_\mu \, (c^{\mu-}_{p'h'})^* \, 
\\  && \nonumber \hspace*{-9.2cm}  
\Bigg[
\langle (\widetilde{\Phi}_p^{\nu +})^*  \widetilde{\Phi}^{\mu-}_{p'}|  
U^{J,{\rm dir}}_{ph,p'h'} | R_h R_{h'} \rangle \, 
- \, \langle (\widetilde{\Phi}_p^{\nu +})^*  \widetilde{\Phi}^{\mu-}_{p'}|  
U^{J,{\rm exc}}_{ph,p'h'} | R_{h'} R_h \rangle 
\Bigg]
\label{eq:eqstr2}
\, .
\eeqn

Putting together Eqs. (\ref{eq:eqstr1}) and (\ref{eq:eqstr2}), we find
a new expression of the CRPA secular equation Eq. (\ref{eq:feq})
\beqn
\hspace*{-0.5cm} 
\sum_\mu \, \sum_{p'h'} \, \Bigg\{ \Bigg[ \delta_{pp'} \, \delta_{hh'}
\Big( \delta_{\mu \nu} \, -
\, \langle (\Phi_p^{\nu +})^*|\cau|\Phi_p^{\mu+} \rangle \,+ \, 
\langle (\Phi_p^{\nu +})^*\, \ide |\caw| \ide \,\Phi^{\mu+}_p \rangle \nonumber \\
&& \hspace*{-7.6cm} + \, \sum_{\epsilon_i<\epsilon_{\rm F}}\, \delta_{ip}\, (\epsilon_i-\epsilon_h-\omega)\,
\langle (\Phi_p^{\nu +})^* |R_i\rangle \langle
(R_i)^*|\Phi_p^{\mu+}\rangle \Big)\nonumber \\
&& \hspace*{-10.cm}  - \, \Big(
\langle (\widetilde{\Phi}_p^{\nu +})^* R_{h'} |  
V^{J,{\rm dir}}_{ph,p'h'} |R_h \widetilde{\Phi}^{\mu+}_{p'} \rangle \,
- \, \langle (\widetilde{\Phi}_p^{\nu +})^* R_{h'} |
   V^{J,{\rm exc}}_{ph,p'h'} | \widetilde{\Phi}^{\mu+}_{p'} R_h  \rangle \Big)
\Bigg] \, c^{\mu+}_{p'h'} \, \nonumber\\
&& \hspace*{-10.cm}  - \, \Big(
\langle (\widetilde{\Phi}_p^{\nu +})^*  \widetilde{\Phi}^{\mu-}_{p'}|  
U^{J,{\rm dir}}_{ph,p'h'} | R_h R_{h'} \rangle \, 
- \, \langle (\widetilde{\Phi}_p^{\nu +})^*  \widetilde{\Phi}^{\mu-}_{p'}|  
U^{J,{\rm exc}}_{ph,p'h'} | R_{h'} R_h \rangle \Big) \, (c^{\mu-}_{p'h'})^*
\Bigg\} \, = \, \nonumber \\
&& \hspace*{-10.5cm}
= \, \langle (\widetilde{\Phi}_p^{\nu +})^* R_{h_0} |
   V^{J,{\rm dir}}_{ph,p_0h_0} |R_h R_{p_0}(\epsilon_{p_0}) \rangle \, 
- \, \langle (\widetilde{\Phi}_p^{\nu +})^* R_{h_0} |
   V^{J,{\rm exc}}_{ph,p_0h_0} |R_{p_0}(\epsilon_{p_0}) R_h  \rangle \, .
\eeqn

A similar equation can be obtained from Eq. (\ref{eq:geq}) for the
$g$ channel function.

\section{Continuum wave function with HF potential}
\label{app:cwf}

We use an expansion on the Sturm-Bessel functions basis to calculate
the s.p. wave function for $\epsilon_p>0$ with HF mean-field
potential.  The explicit expression of the differential equation to be
solved for the reduced radial part of the wave function
$u_p(r)=r\,R_p(r,\epsilon_p)$ is
\beq 
-\, \frac{\hbar^2}{2m} \, \left( \frac{{\rm d}^2}{{\rm d}r^2} \, 
- \, \frac{l_p(l_p+1)}{r^2} \right) \, u_p(r) \, + \, 
\cau(r) \, u_p(r) \, - \, \int {\rm d}r' \, r' \, 
\caw(r,r')\, u_p(r') \, = \, \epsilon_p \,u_p(r) \, , 
\eeq 
where $\cau$ and $\caw$ have been defined in
Eqs. (\ref{eq:hartree}) and (\ref{eq:dirac}) respectively.

We express the solution of the above equation as 
\beq 
\frac{u_p(r)}{r}\, = \, j_{l_p}(k_pr) \, + \, \sum_\mu \, c^\mu_p \, 
\Phi^\mu_p (r) 
\eeq 
where $j_{l_p}(k_pr)$ is a spherical Bessel function with $k_p$ the
wave number corresponding to $\epsilon_p$. Using the definition
(\ref{eq:sturmb}) of the Sturm-Bessel functions, and their
orthogonality relation (\ref{eq:ort1}), we obtain the following non
homogeneous system
\beqn 
\sum_\mu \Bigg[ \delta_{\mu\nu} \, - \, \int {\rm d}r \, r^2 \,
\Phi^\nu_p(r) \, \cau(r) \, \Phi^\mu_p(r) \, + \, \int {\rm d}r \, r^2
\, \Phi ^\nu _p (r) \, \int {\rm d}r' \, r'^2 \, \caw(r,r') \, \Phi
^\mu_p(r') \Bigg] \, c^\mu_p \, = \nonumber && \\ &&\hspace*{-14cm} = \, 
\int {\rm d}r \, r^2 \, \Phi^\nu_p (r) \, \cau(r) \, j_{l_p}(k_p r) \,  - \, 
\int {\rm d}r \, r^2 \, \Phi^\nu_p(r) \, \int {\rm d}r' \, r'^2 \, 
\caw(r,r') \, j_{l_p}(k_p r') \, , 
\eeqn
where the unknowns are the expansion coefficients $c^\mu_p $.
%
%
\acknowledgments We thank I. Sick for providing us with elastic
electron scattering data.  This work has been partially supported by
the Spanish Ministerio de Ciencia e Innovaci\'on under contracts
FPA2009-14091-C02-02 and ACI2009-1007 and by the Junta de
Andaluc\'{\i}a (FQM0220).

\begin{thebibliography}{10}
\expandafter\ifx\csname url\endcsname\relax
  \def\url#1{\texttt{#1}}\fi
\expandafter\ifx\csname urlprefix\endcsname\relax\def\urlprefix{URL }\fi

\bibitem{wir95}
R.~B. Wiringa, V.~G.~J. Stoks, R.~Schiavilla, Phys.\ Rev. \ C 51 (1995) 38.

\bibitem{mac01}
R.~Machleidt, Phys. \ Rev. \ C 63 (2001) 024001.

\bibitem{pud97}
B.~S. Pudliner, V.~R. Pandharipande, J.~Carlson, S.~C. Pieper, R.~B. Wiringa,
  Phys.\ Rev. \ C 56 (1997) 1720.

\bibitem{pie02}
S.~C. Pieper, K.~Varga, R.~B. Wiringa, Phys. \ Rev. \ C 66 (2002) 044310.

\bibitem{pie01}
S.~C. Pieper, R.~B. Wiringa, Ann.\ Rev. \ Nucl. \ Part. \ Sc. 51 (2001) 53.

\bibitem{kam01}
H.~Kamada, et~al., Phys.\ Rev. \ C 64 (2001) 044001.

\bibitem{nav00}
P.~Navr\'atil, J.~P. Vary, B.~R. Barrett, Phys.\ Rev. \ C 62 (2000) 054311.

\bibitem{dea04}
D.~J. Dean, M.~Hjorth-Jensen, Phys. \ Rev. \ C 69 (2004) 054320.

\bibitem{gan06}
S.~Gandolfi, F.~Pederiva, S.~Fantoni, K.~E. Schmidt, Phys.\ Rev. \ C 73 (2006)
  044304.

\bibitem{ari07}
F.~Arias~de Saavedra, C.~Bisconti, G.~Co', A.~Fabrocini, Phys. \ Rep. 450
  (2007) 1.

\bibitem{qua09}
S.~Quaglioni, P.~Navr{\'a}til, Phys. \ Rev. \ C 79 (2009) 044606.

\bibitem{rot10}
R.~Roth, T.~Neff, H.~Feldmeier, Prog. \ Part. \ Nucl. \ Phys. 65 (2010) 50.

\bibitem{rin80}
P.~Ring, P.~Schuck, The nuclear many-body problem, Springer, Berlin, 1980.

\bibitem{spe77}
J.~Speth, E.~Werner, W.~Wild, Phys. \ Rep. 33 (1977) 127.

\bibitem{spe80}
J.~Speth, V.~Klemt, J.~Wambach, G.~E. Brown, Nucl. \ Phys. \ A 343 (1980) 382.

\bibitem{mig67}
A.~Migdal, Theory of finite Fermi systems and applications to atomic nuclei,
  Interscience, London, 1967.

\bibitem{ber75}
G.~F. Bertsch, S.~F. Tsai, Phys. \ Rep. 18 (1975) 125.

\bibitem{shl75}
S.~Shlomo, G.~F. Bertsch, Nucl. \ Phys. \ A 243 (1975) 507.

\bibitem{liu76}
K.~F. Liu, N.~Van Giai, Phys. \ Lett. \ B 55 (1976) 23.

\bibitem{van81}
N.~Van Giai, H.~Sagawa, Nucl. \ Phys. \ A 371 (1981) 1.

\bibitem{war87}
M.~Waroquier, et~al., Phys. \ Rep. 148 (1987) 249.

\bibitem{sar93}
A.~M. Saruis, Phys. \ Rep. 235 (1993) 57.

\bibitem{ham96}
I.~Hamamoto, H.~Sagawa, X.~Z. Zhang, Phys. \ Rev. \ C 53 (1996) 765.

\bibitem{dec80}
J.~Decharg\`e, D.~Gogny, Phys. \ Rev. \ C 21 (1980) 1568.

\bibitem{mar07}
M.~Martini, G.~Co', M.~Anguiano, A.~M. Lallena, Phys.\ Rev. \ C 75 (2007)
  034604.

\bibitem{co90}
G.~Co', A.~M. Lallena, Nucl. \ Phys. \ A 510 (1990) 139.

\bibitem{don09}
V.~De~Donno, G.~Co', C.~Maieron, M.~Anguiano, A.~M. Lallena, M.~Moreno-Torres,
  Phys. \ Rev. \ C 79 (2009) 044311.

\bibitem{uda89}
T.~Udagawa, B.~T. Kim, Phys. \ Rev. \ C 40 (1989) 2271.

\bibitem{bub91}
M.~Buballa, S.~Dro\.zd\.z, S.~Krewald, J.~Speth, Ann. \ of \ Phys. 208 (1991)
  346.

\bibitem{nak09}
H.~Nakada, K.~Mizuyama, M.~Yamagami, M.~Matsuo, Nucl. \ Phys. A 828 (2009) 283.

\bibitem{piek01}
J.~Piekarewicz, Phys. \ Rev. \ C 64 (2001) 024307.

\bibitem{boh81}
O.~Bohigas, N.~Van Giai, D.~Vautherin, Phys. \ Lett. \ B 102 (1981) 105.

\bibitem{edm57}
A.~R. Edmonds, Angular momentum in quantum mechanics, Princeton University
  Press, Princeton, 1957.

\bibitem{rot62}
M.~Rotenberg, Ann. \ Phys. 19 (1962) 262.

\bibitem{wei63}
S.~Weinberg, Phys. \ Rev. \ B. 133 (1963) 232.

\bibitem{wei64}
S.~Weinberg, Lectures on Particles and Field Theory, Brandeis Summer Institute
  in Theoretical Physics, {\rm S. Deser and K. W. Ford Eds.}, Prentice-Hall,
  Egelwood Cliffs, 1964.

\bibitem{new66}
G.~Newton, Scattering theory of Waves and Particles, McGraw-Hill, New York,
  1966.

\bibitem{raw82}
G.~Rawitscher, Phys. \ Rev. \ C. 25 (1982) 2196.

\bibitem{eisII}
J.~M. Eisenberg, W.~Greiner, Excitation Mechanisms of the nucleus, North
  Holland, Amsterdam, 1970.

\bibitem{bla52}
J.~M. Blatt, V.~F. Weisskopf, Theoretical nuclear physics, John Wiley and sons,
  New York, 1952.

\bibitem{ber91}
J.~F. Berger, M.~Girod, D.~Gogny, Comp. \ Phys. \ Commun. 63 (1991) 365.

\bibitem{gor09}
S.~Goriely, S.~Hilaire, M.~Girod, S.~P\'eru, Phys. \ Rev. \ Lett. 102 (2009)
  242501.

\bibitem{akm98}
A.~Akmal, V.~R. Pandharipande, D.~G. Ravenhall, Phys. \ Rev. \ C 58 (1998)
  1804.

\bibitem{gan10}
S.~Gandolfi, A.~Y. Illarionov, S.~Fantoni, J.~C. Miller, F.~Pederiva, K.~E.
  Schmidt, Mont. \ Not. \ R. \ Astron. \ Soc. 404 (2010) L35.

\bibitem{co09b}
G.~Co', V.~De~Donno, C.~Maieron, M.~Anguiano, A.~M. Lallena, Phys. \ Rev. \ C
  80 (2009) 014308.

\bibitem{dej87}
C.~W.~De Jager, C.~De Vries, At. \ Data \ Nucl. \ Data \ Tables 36 (1987) 495.

\bibitem{ann95a}
R.~Anni, G.~Co', P.~Pellegrino, Nucl. \ Phys. \ A 584 (1995) 35.

\bibitem{sic70}
I.~Sick, J.~McCarthy, Nucl. \ Phys. \ A 150 (1970) 631.

\bibitem{sic}
I.~Sick, private communication.

\bibitem{dao09}
J.~Daoutidis, P.~Ring, Phys. \ Rev. \ C 80 (2009) 024309.

\bibitem{kno75}
K.~T. Kn{\"o}pfle, G.~J. Wagner, H.~Breuer, M.~Rogge, C.~Mayer-B{\"o}ricke,
  Phys. \ Rev. \ Lett. 35 (1975) 779.

\bibitem{ahr75}
J.~Ahrens, et~al., Nucl. \ Phys. \ A 251 (1975) 479.

\bibitem{tra87}
M.~Traini, G.~Orlandini, R.~Leonardi, Rivista \ Nuovo \ Cimento 10 (1987) 1.

\bibitem{deh82}
R.~de~Haro, S.~Krewald, J.~Speth, Nucl. \ Phys. \ A 388 (1982) 265.

\bibitem{co85}
G.~Co', S.~Krewald, Nucl. \ Phys. \ A 433 (1985) 392.

\bibitem{sin73}
B.~B.~P. Sinha, G.~Peterson, R.~R. Whitney, I.~Sick, J.~McCarthy, Phys. \ Rev.
  \ C 7 (1973) 1930.

\bibitem{sic79}
I.~Sick, J.~Bellicard, J.~Cavedon, B.~Frois, M.~Huet, P.~Leconte, P.~Ho,
  S.~Platchkov, Phys. \ Lett. \ B 88 (1979) 240.

\bibitem{cav80t}
J.~M. Cavedon, Ph.D. thesis, Universit\'e de Paris-Sud (France), unpublished
  (1980).

\bibitem{sic75a}
I.~Sick, Model-independent densities of s/d-shell nuclei, unpublished (1975).

\bibitem{lui81}
Y.-W. Lui, J.~D. Bronson, C.~M. Rozsa, D.~H. Youngblood, P.~Bogucki, U.~Garg,
  Phys. \ Rev. \ C 24 (1981) 884.

\bibitem{ber84}
I.~Bergqvist, R.~Zorro, A.~H{\aa}kansson, A.~Lindholm, L.~Nilsson, N.~Olsson,
  A.~Likar, Nucl. \ Phys. \ A 419 (1984) 509.

\bibitem{dro90}
S.~Dro\.zd\.z, S.~Nishizaki, J.~Speth, J.~Wambach, Phys. \ Rep. 197 (1990) 1.

\bibitem{kam04}
S.~Kamerdzhiev, J.~Speth, G.~Tertychny, Phys. \ Rep. 393 (2004) 1.

\bibitem{gam10}
D.~Gambacurta, M.~Grasso, F.~Catara, Phys. \ Rev. \ C 81 (2010) 054312.

\bibitem{co10}
G.~Co', V.~De~Donno, M.~Anguiano, A.~M. Lallena, arxiv:1009.3364 [nucl-th].

\bibitem{ots06}
T.~Otsuka, T.~Matsuo, D.~Abe, Phys. \ Rev. \ Lett. 97 (2006) 162501.

\bibitem{mor10}
M.~Moreno-Torres, M.~Grasso, H.~Liang, V.~de~Donno, M.~Anguiano, N.~Van Giai,
  Phys. \ Rev. \ C 81 (2010) 064327.

\end{thebibliography}

\end{document}